\documentclass{rcdj}

\usepackage{hhline}

\begin{document}

\setcounter{secnumdepth}{3}
\setcounter{tocdepth}{5}
\makeatletter
\renewcommand*\l@paragraph{\@dottedtocline{4}{1.0em}{2.1em}}
\@addtoreset{paragraph}{section}
\makeatother

\setcounter{page}{177}
\journal{REGULAR AND CHAOTIC DYNAMICS, V.\,7, \No2, 2002}
\title{THE ROLLING MOTION OF A RIGID BODY ON A PLANE\\ AND A SPHERE.
HIERARCHY OF DYNAMICS}
\runningtitle{THE ROLLING MOTION OF A RIGID BODY ON A PLANE AND A SPHERE.
HIERARCHY OF DYNAMICS}
\runningauthor{A.\,V.\,BORISOV, I.\,S.\,MAMAEV}
\authors{A.\,V.\,BORISOV}
{Department of Theoretical Mechanics\\
Moscow State University,
Vorob'ievy Gory\\
119899, Moscow, Russia\\
E-mail: borisov@rcd.ru}
\authors{I.\,S.\,MAMAEV}
{Laboratory of Dynamical Chaos and Nonlinearity\\
Udmurt State University, Universitetskaya, 1\\
426034, Izhevsk, Russia\\
E-mail: mamaev@rcd.ru}
\amsmsc{37J60, 37J35}
\received 17.01.2002.
\doi{10.1070/RD2002v007n02ABEH000204}
\abstract{In this paper we consider cases of existence of invariant measure,
additional first integrals, and Poisson structure in a problem of rigid
body's rolling without sliding on plane and sphere. The problem of rigid
body's motion on plane was studied by S.\,A.\,Chaplygin, P.\,Appel,
D.\,Korteweg. They showed that the equations of motion are reduced to a
second-order linear differential equation in the case when the surface of
dynamically symmetric body is a surface of revolution. These results were
partially generalized by P.\,Woronetz, who studied the motion of body
of revolution and the motion of round disk with sharp edge on the surface of
sphere. In both cases the systems are Euler\f Jacobi integrable and have
additional integrals and invariant measure. It turns out that after some
change of time defined by reducing multiplier, the reduced system is a
Hamiltonian system. Here we consider different cases when the integrals
and invariant measure can be presented as finite algebraic expressions.

We also consider the generalized problem of rolling of dynamically
nonsymmetric Chaplygin ball. The results of studies are presented as tables
that describe the hierarchy of existence of various tensor invariants:
invariant measure, integrals, and Poisson structure in the considered
problems.}

\maketitle

\section*{Contents}

\makeatletter
\@starttoc{toc}
\makeatother

\bigskip

\section{Equations of Rigid Body Motion on Plane and Sphere without
Sliding (Nonholonomic Rolling)}

\wfig<bb=0 0 56.8mm 41.8mm>{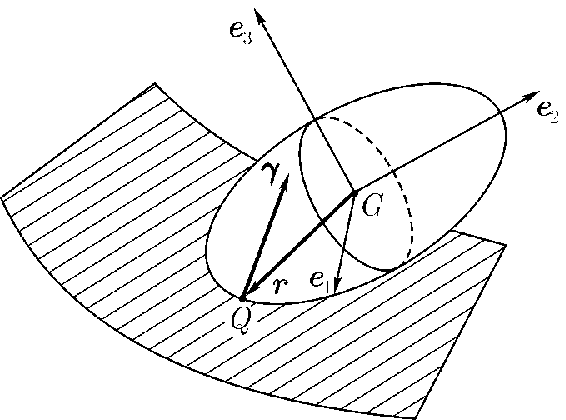}

In this paper we consider the equations of rigid body's rolling on plane
and sphere, because in these two cases, as opposed to a rolling on an
arbitrary surface, the equations of motion are similar the Euler\f Poisson
equations. In both cases there are six first-order equations for six
variables. In the potential field the equations have two integrals of
motion: the energy integral and the geometrical integral (for Euler\f
Poisson equations there is also the area integral; its analog in problem
of rolling is presented only under additional dynamical and geometrical
restrictions).

Suppose that the rigid body rolls without sliding (i.e. the velocity of
contact point~$Q$ is equal to zero) on the fixed surface represented by
plane or sphere. The first part of equations of motion is the vector
dynamical equation of kinetic moment~$\bM$ behavior in time with respect
to the contact point~$Q$ (Fig.~\ref{ris1.eps}). This equation is
represented for \emph{arbitrary shapes of body and surface} in the form
\eq[eq1.1]{
\dot{\bs M}={\bs M}\times{\bs \omega}+m\dot{\bs r}\times
({\bs\omega}\times{\bs r})+{\bs M}_Q,
}
where~${\bs M},\,{\bs \omega},\,{\bs r}={GQ}$, $\bM_Q$ are supposed to be
projected on the principal central axes of inertia in the body; here
${\bs \omega}$ is the angular velocity, ${\bs M_Q}$ is the moment of
external forces with respect to the contact point, $G$ is the center of
mass. The second part of the motion equation is the vector kinetic
equation of Poisson type different for the cases of plane~{\bf a)}
and sphere~{\bf b)}:
\begin{itemize}
\item[{\bf a)}]\mbox{}\vspace{-6mm}
\eq[eq1.2]{
\dot{\bs\gam}={\bs\gam}\times{\bs\omega},
}
where~${\bs\gam}$ is the unit vector orthogonal to the plane,
\item[{\bf b)}]\mbox{}\vspace{-6mm}
\eq[eq1.3]{
{R}_0(\dot{\bs\gam} + {\bs\omega} \times
{\bs\gam}) =\dot{\bs r},
}
where~${\bs\gam}$ is the unit vector orthogonal to the sphere of
radius~$R_0$ (see Fig.~\ref{ris2.eps}).
\end{itemize}

In equations~(\ref{eq1.1}),\,(\ref{eq1.2}),\,(\ref{eq1.3}) we suppose
that the radius vector~${\bs r}$ is expressed as a function of normal
vector~$\gam$ with the help of equation
\eq[eq1.4]{
{\bs\gam}=-\frac{\grad{f}}{|\grad{f}|}
}
that define the Gauss transformation, where~$f(\bs r)=0$ is the equation
of the body's surface in the main central frame of references connected
with the body. We suppose that the body is everywhere convex (to exclude
the collisions during the motion), and equation~(\ref{eq1.4}) is uniquely
solvable with respect to~${\bs r}=\br({\bs\gam})$. We assume that this
condition is always fulfilled in the following text.

Using~(\ref{eq1.4}) we can represent kinematic equation~(\ref{eq1.3})
describing the dynamics of vector~${\bs\gam}$ in the case of sphere in the
form
\eq[eq1.5]{
\dot{\bs\gam}=\left(1+k({\bf B}-k)^{-1}\right) {\bs\gam}\times{\bs\om},
\qq k=R_0^{-1},
}
where~${\bf B}=\|b_{ij}\|$ is a degenerate matrix with the components
$b_{ij}=-\pt{}{r_i}\Bigl(\frac{1}{|\nabla f|}\pt{f}{r_j}\Bigr)$.

The relation between~$\bM$ and~$\bs\omega$ is defined by the equation
\eq[eq1.6]{
{\bs M}= {\bf I}{\bs \omega}+m{\bs r}\times ({\bs\omega}\times{\bs
r}),\quad
\bs\omega=\frac{{\bf A}\bM-m\br\times({\bf A}\br\times{\bf A}\bM)}{1-m(\br,{\bf
A}\br)},
}
where~$m$ is the mass of body, ${\bf I}$ is the central tensor of inertia,
${\bf A}=({\bf I}+m\br^2)^{-1}$.\goodbreak

If the potential~$U=U(\bs\gamma)$ depends only on the components of
vector~$\bs\gamma$, then we can present the moment of external
forces in the form:

{\bf a)} in the case of plane~${\bs M}_Q={\bs
\gam}\times\pt{U}{{\bs \gam}}$;

{\bf b)} in the case of sphere~$ {\bs M}_Q={\bs
\gam}\times(1+k({\bf B}^{\rm T}-k)^{-1})\pt{U}{\bs \gam}$,
where~$\bf B$ is matrix~(\ref{eq1.5}).

For equations~(\ref{eq1.1}) and~(\ref{eq1.2}),(\ref{eq1.3}) we always have
the energy integral and the geometrical integral
\eq[eq1.7]{
H=\frac{1}{2}({\bs M},\,{\bs\om})+U({\bs\gam}), \quad
F_1=\bs\gamma^2=1.
}

\begin{rem*}
The proof is based on the formula
$$
\frac12(\bM,\bs\omega)^{\ds\bs.}=(\bM_Q,\bs\omega).
$$
It follows only from equation~\eqref{eq1.1} and does not depend on the
shape of the surface, on which the body is rolling.
\end{rem*}

According to Euler\f Jacobi theorem (the theory of last multiplier), %??????
to integrate these equations we need two more independent first integrals
and an invariant measure~\cite{kozlov1}. Recall that the density of
invariant measure~$\rho$ of the general system
$$
\dot{\bx}=\bv(\bx),\quad \bx =(x_1,\ldots,x_n)
$$
satisfies the Liouville equation~$\div(\rho\bv)=0$. In the general case
none of these objects exists; therefore, the system shows some
interesting asymptotic and chaotic properties characteristic for the
oscillation of Celtic stones. %?????????
Such properties are not typical for conservative
systems~\cite{karapet,kozlov1}. We consider all the known cases of
existence of additional first integrals (one or two at once) and the
cases of existence of invariant measure.

Dynamical and geometrical bounds leading to the existence of the first
integrals and of the invariant measure are independent in some sense. For
some combinations of parameters there exist only the measure or only the
additional integral. In some extreme cases two additional integrals and
measure can exist at the same time; thus, the system became completely
integrable.

The results of the study are presented separately for the cases of body's
rolling on plane and on sphere and collected in tables~1, 2. The following
subsections are essentially the comments for these tables.

\section{Body on a Plane}

\paragraph{Body of revolution's rolling on plane
(S.\,A.\,Chaplygin~\cite{chaplygin2}, P.\,Appell~\cite{Appell,appel-eng2}).}
If both the surface of body and the central ellipsoid of inertia are
coaxial surfaces of revolution, then for equations~\eqref{eq1.1},
\eqref{eq1.2} there exist two additional integrals and invariant measure.
We assume that the potential~$U$ is an arbitrary function
of~$\gamma_3=\cos\theta$, i.\,e. it depends only on the slope of revolution
axis of the body to the vertical. In particular, in the case of the gravity
field the center of mass must be situated on the axis of revolution.

The integrability of problem for an arbitrary body of revolution was shown
by S.\,A.\,Chaplygin in 1897~\cite{chaplygin2}. He also demonstrated that
it is possible to add a balanced uniformly revolving rotor along the axis
of revolution (gyrostat) preserving the integrability of problem. More
specific cases of this problem were studied by

a. Routh (1884): the rolling of unbalanced dynamically symmetric ball
on plane.

b. Neumann, Carvallo (1898), Appell (1899) and Korteweg (1900): the
rolling of round disk.

The results of Neumann and Carvallo mainly concern the deduction of motion
equations and determination of stationary solutions. Note that Neumann
during the deduction of equations of motion at first made the same mistake
that occurred before in Lindel\"{o}f paper. He applied the Lagrangean
formalism without the necessary "nonholonomic"\ modifications. In the
subsequent studies he corrected this mistake, but did not solve the
problem in quadratures. The lindel\"{o}f mistake was analyzed in detail
by S.\,A.\,Chaplygin (1897). He obtained a new form of equations of
nonholonomic dynamics and was able to reduce the considered problem of
rolling of revolution body to two linear first-order equations. In the
case of round disk's rolling, S.\,A.\,Chaplygin showed the possibility of
reduction of these two equations to one linear (second-order) equation
solvable in hypergeometric functions. We should also note that, before
Chaplygin's work, the equations of heavy revolution body motion were
obtained in~1861 by G.\,Slesser, but their integrability was not indicated.

Somewhat later (in 1898), the analogous substitution (in the equations
obtained by Carvallo in the paper presented for the Fourneyron prize) was
used by Appell and in the slightly different form by Korteweg. They both
did not know S.\,A.\,Chaplygin's paper that was published in inaccessible
journal only in Russian (the English translation of Chaplygin's papers
dedicated to this problem is published in this journal in 2002). This is
the reason of the fact that in many modern textbooks and research papers
(O'Reily~\cite{oreilly}) the problem of round disk's rolling is connected
with the names of Appell and Korteweg, although the previous text show
that this opinion is not completely correct.

Here we present the results obtained by S.\,A.\,Chaplygin in the modern
algebraic form that let us to show the invariant measure in the explicit
form and also to obtain the simplest forms of the first integrals. It
turns out that we can generalize these results to dynamically
nonsymmetric situation.

In the case of body of revolution we can find the solutions of equation
of surface~(\ref{eq1.6}) in the explicit form
\eq[eq2.1]{
r_1=f_1(\gam_3)\gam_1, \q r_2=f_1(\gam_3)\gam_2, \q
r_3=f_2(\gam_3),
}
where~$f_i(\gam_3)$, $i=1,2$ are function subjected to the differential
equation that defines the meridional section
\begin{equation}
\label{eq2.1/2}
\frac{df_2}{d\gamma_3}=f_1-\frac{1-\gamma_3^2}{\gamma_3}\frac{df_1}{d\gamma_3}.
\end{equation}

If we denote the main central tensor of inertia
as~${\bf I}=\diag(I_1,\,I_1,\,I_3)$, $(I_1=I_2)$, then we can explicitly
calculate the density of invariant measure of
equations~(\ref{eq1.1}),\,(\ref{eq1.2}). It exists for arbitrary
functions $f_1(\gamma_3)$, $f_2(\gamma_3)$ that define the surface
\eq[eq2.2]{
\rho=\frac{1}{\sqrt{I_1I_3+m(\br,{\bf I}\br)}}
=\frac{1}{\sqrt{I_1I_3+I_1mf_1^2(1-\gam_3^2)+I_3mf_2^2}}.
}

\begin{rem*}
For the equations motion in variables $\bs\omega,\bs\gamma$ the density of
invariant measure differs from~\eqref{eq2.2} by factor $\det {\bf I}_Q$,
where ${\bf I}_Q={\bf I}+m(\br^2{\bf E}-\br\otimes\br)$ is the tensor of
inertia with respect to the point of contact. In the case $I_1=I_2$ we
have $\det{\bf I}_Q=(I_1+m\br^2)(I_1I_3+m(\br,{\bf I}\br))$. It is
interesting that measure~\eqref{eq2.2} contain one of these factors.
\end{rem*}

It is easy to verify that under the above conditions the equations of
motion also have {\em the symmetry field\/}~${\bs v}$ defined by the
differential operator
\eq[eq2.3]{
\wh {\bs v}=M_1\pt{}{M_2}-M_2\pt{}{M_1}+
\gam_1\pt{}{\gam_2}-\gam_2\pt{}{\gam_1}.
}
It corresponds to the invariance of the system with respect to rotations
about the axis of dynamical symmetry. Using this field we can reduce the
order of system. For that we should choose the integrals of vector
fields~\eqref{eq2.3} as reduced variables to present the equations in
the simplest form. After a number of tries, we choose the following
reduced variables
\eq[eq2.4]{
\begin{gathered}
K_1=\frac{(\bM,\br)}{f_1}=M_1\gamma_1+M_2\gamma_2+\frac{f_2}{f_1}M_3\\
K_2=\frac{\omega_3}{\rho}=\rho\Bigl(mf_1f_2(M_1\gamma_1+M_2\gamma_2)+
(I_1+mf_2^2)M_3\Bigr)\\
K_3=\frac{M_2\gamma_1-M_1\gamma_2}{\sqrt{(1-\gamma_3^2)(I_1+m\br^2)}}.
\end{gathered}
}
In these variables the equations of motion of the reduced system
have the following form
\eq[eq2-*1]{
\begin{gathered}
\dot\gamma_3=kK_3\\
\dot K_1=-kK_3\rho I_3\Bigl(1-\Bigl(\frac{f_2}{f_1}\Bigr)'\Bigr)K_2,\\
\dot K_2=-kK_3\rho mf_1\bigl(f_1-f_2'\bigr)K_1,\\
\dot K_3=-\frac{k}{I_1^2f_1^2(1-\gamma_3^2)^2}\biggl(
f_2\bigl(f_1(1-\gamma_3^2)+\gamma_3f_2\bigr)(mf_1^2K_1^2+I_3K_2^2)+\\
+\gamma_3f_1^2I_1K_1^2-f_1\bigl(f_1(1-\gamma_3^2)+2\gamma_3f_2\bigr)
\frac{K_1K_2}{\rho}+\\
+mf_1^2\rho f_2(1-\gamma_3^2)(\gamma_3f_1I_1-f_2I_3)K_1K_2\biggr)-
k\pt{U(\gamma_3)}{\gamma_3},
\end{gathered}
}
where~$k=\sqrt{\frac{1-\gamma_3^2}{I_1+m\br^2}}$.

It is easy to show that these equations have the invariant measure with
density~$\rho=k^{-1}$ and integral of energy
\begin{equation}
\label{eq2-*2}
\begin{gathered}
H=\frac12(\bM,\bs\omega)+U(\gamma_3)=\\
=\frac{1}{2I_1(1-\gamma_3^2)}\Bigl(K_1^2-\frac{I_3}{mf_1^2}K_2^2+\frac{mf_2^2}{I_1}
\Bigl(K_1-\frac{K_2}{\rho m
f_1f_2}\Bigr)^2\Bigr)+\frac{1}{2}K_3^2+U(\gamma_3).
\end{gathered}
\end{equation}
Moreover, for system~\eqref{eq2-*1} we have the following theorem

\begin{teo*}
After the change of time $k\,dt=d\tau$ vector field~\eqref{eq2-*1} become
Hamiltonian
$$
\frac{dx_i}{d\tau}=\{x_i,H\},\quad \bx=(\gamma_3,K_1,K_2,K_3)
$$
with Hamiltonian~\eqref{eq2-*2} and degenerate Poisson bracket:
\begin{equation}
\label{eq2-*3}
\begin{gathered}
\{\gamma_3,K_3\}=1,\quad
\{K_1,K_3\}=-I_3\rho\Bigl(1-\Bigl(\frac{f_2}{f_1}\Bigr)'\Bigr)K_2,\\
\{K_2,K_3\}=-m\rho f_1(f_1-f_2')K_1
\end{gathered}
\end{equation}
{\rm (}all the other brackets are equal to zero\/{\rm)}.
\end{teo*}

The proof of theorem is obtained by the direct verification of
equations and of Jacobi identity.

It turns out that equation~\eqref{eq2-*1} can be written in antisymmetric,
almost Hamiltonian form (that sometimes is referred to as antigradient form)
\begin{equation}
\label{eq2a-*1}
\frac{dx_i}{d\tau}={J_\lambda}_{ij}\pt{H}{x_j},\quad {\bf J}_\lambda=
-{\bf J}_\lambda^{\text{T}},
\end{equation}
where
\begin{equation}
\label{eq2a-*2}
\begin{gathered}
{\bf J}_\lambda=\begin{pmatrix}
0 & 0 & 0 & 1\\
0 & 0 & \lambda & -I_3\rho\Bigl(1-\Bigl(\frac{f_2}{f_1}\Bigr)'\Bigr)
K_2-\lambda u\\
0 & -\lambda & 0 & -m\rho f_1(f_1-f_2')K_1-\lambda v\\
-1 & I_3\rho \Bigl(1-\Bigl(\frac{f_2}{f_1}\Bigr)'\Bigr)K_2+\lambda u &
m\rho f_1(f_1-f_2')K_1+\lambda v & 0
\end{pmatrix}\\
u=\frac{1}{f_1^2I_1^2(1-\gamma_3^2)K_3}\Bigl(\frac{(\br,{\bf I}\br)}{m}
K_2-\frac{f_1f_2}{\rho}K_1\Bigr),\\
v=\frac{1}{I_1^2(1-\gamma_3^2)K_3}\Bigl(\frac{f_2}{f_1\rho}K_2-
(I_1+mf_2^2)K_1\Bigr),
\end{gathered}
\end{equation}
and $\lambda$ is an arbitrary function of~$(K_1,K_2,K_3,\gamma_3)$.
At $\lambda=0$ we obtain the degenerate tensor ${\bf J}_0$ corresponding
to bracket~\eqref{eq2-*3}; although for $\lambda\ne 0$
tensor~\eqref{eq2a-*2} is nondegenerate, it does not satisfy the Jacobi
identity, i.\,e. it does not define a Poisson bracket.

If $\lambda$ is chosen in the form
\begin{equation}
\label{eq2a-*3}
\lambda=\alpha (\gamma_3)K_3,
\end{equation}
then the tensor $\wt{\bf J}=(\lambda^{-1}{\bf J}_\lambda)$ satisfies
the Jacobi identity, and the corresponding vector field
\begin{equation}
\label{eq2a-*4}
\bv=(\lambda^{-1}{\bf J}_\lambda)\nabla H
\end{equation}
is Hamiltonian; at the same time the divergence of field~\eqref{eq2a-*4}
is nonzero. Thus, the considered nonholonomic system generate an example
of Hamiltonian vector field with nontrivial measure $\rho=\alpha
(\gamma_3)K_3$. Note also that the function $\rho=\alpha (\gamma_3)K_3$
is a reducing multiplier in Chaplygin's terminology, and in this case it
differs from invariant measure~\eqref{eq2.2}. The close example of Poisson
structure for the problem of ball's rolling on body of revolution was
presented by Hermans~\cite{hermans}.

Bracket \eqref{eq2-*3} has two Casimir functions~\cite{bormam-2} that are
integrals of motion; therefore, system~\eqref{eq2-*1} is integrable.
The integrability and existence of linear integrals can be established by
the different classical method: we divide the second and the third
equation of system~\eqref{eq2-*1} by $\dot\gamma_3$ and obtain the system
of two linear non-autonomous first-order equations
\begin{equation}
\label{eq2.5}
\frac{dK_1}{d\gamma_3}=-\rho
I_3\Bigl(1-\Bigl(\frac{f_2}{f_1}\Bigr)'\Bigr)K_2,\quad
\frac{dK_2}{d\gamma_3}=-m\rho  f_1(f_1-f_2')K_1.
\end{equation}
In somewhat different variables connected with semifixed axes,
equations~(\ref{eq2.5}) were obtained by S.\,A.\,Chaplygin.
Equations~(\ref{eq2.5}) do not contain the potential, which is presented
only in the expression of energy integral~\eqref{eq2-*2}. Using this
equation we determine the dependence of nutation angle on time after
the solving of linear system~(\ref{eq2.5}). In the general case this
dependence has periodic oscillating character.

Because the equations~(\ref{eq2.5}) are linear with respect to~$\gam_3$,
the general solution can be obtained as the linear superposition
\eq[eq2.7]{
K_i=c_1g_i^{(1)}+c_2g_i^{(2)},\q i=1,2,
}
where~$g_i^{(1)}(\gam_3),\,g_i^{(2)}(\gam_3)$ are elementary solutions
of~\eqref{eq2.5}. Inverting expressions~(\ref{eq2.7}) with respect
to~$c_i$ we obtain the expressions for the lacking first integrals, which
are expressed in the general case in terms of real analytic, but
nonalgebraic (for example, hypergeometric) functions. Nevertheless,
they are always linear with respect to~$M_i$ (i.\,e. with respect to
generalized velocities).
%????

These integrals in some sense generalize the area integral (corresponding
to the cyclic angle of precession) and the cyclic Lagrange integral
(corresponding to the cyclic angle of proper rotation)~\cite{bormam-2}
that exist in the classical problem of heavy symmetric top's motion
about a fixed point (the Lagrange case). The presence of such integrals
causes the great similarity of qualitative researches of these
problems.

Let's consider all known situations when integrals are algebraic or can be
expressed through members of some known classes of special functions.

\paragraph{Round disk (S.\,A.\,Chaplygin, P.\,Appell, D.\,Korteweg).}
Generally speaking, we consider a disk with the center of mass
displaced along the axis of dynamical symmetry (Fig.~\ref{ris5.eps}).
In this case functions~(\ref {eq2.1}) are explicitly expressed as
\eq[eq2.71]{
f_1=\frac{R}{\sqrt{1-\gam^2_3}}, \qq f_2=a,
}
where~$R$ is a radius of coin, $a$ is the displacement of the center of
mass along the axis of dynamical symmetry (Fig.~\ref{ris5.eps}).

\wfig<bb=0 0 48.2mm 22.9mm>{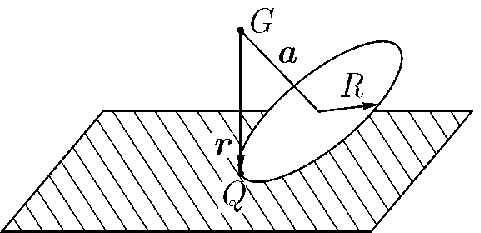}

The interesting fact in this case is the independence of
measure~\eqref{eq2.2} from the phase variables~$\rho=\const$. For
variables~\eqref{eq2.4} we obtain the equations
\eq[eq2.8]{
\frac{dK_1}{d\theta}=\frac{\rho mR^2}{\sin\theta}K_2, \qq
\frac{dK_2}{d\theta}=I_3\rho(\sin\theta+\frac{a}{R}\cos\theta)K_1,
}
where~$\rho=(I_1I_3+I_1mR^2+I_3ma^2)^{-\fracs12}$. These two linear
equations are reduced to one linear second-order equation with respect
to~$\om_3$
\eq[eq2.9]{
\frac{d^2\omega_3}{d\theta^2}-\ctg\theta\frac{d\omega_3}{d\theta}+\rho^2mR(R+
a\ctg\theta)I_3\omega_3=0.
}
At $a=0$ with the help of substitution $\cos\theta=1-2x$
equation~\eqref{eq2.9} is transformed to the hypergeometric type
equation~\cite{Appell}
\eq*{
x(1-x)\frac{d^2\omega_3}{dx^2}+(1-2x)\frac{d\omega_3}{dx}-\rho^2I_3mR^2\omega_3=0.
}

In the papers~\cite{kolesnik2,fedorov} the following result was shown:
the disk for almost all initial conditions do not fall on the plane.
In~\cite{kozaf} the similar result was obtained for the nonintegrable
problem of heavy disk's rolling on a slopping plane.

The results concerning the stability of stationary motions and the
qualitative analysis of motion see in the papers~\cite{kuleshov1,fedorov}
and also in the book~\cite{markeev1}.

\paragraph{Dynamically symmetric ball with the displaced center
of mass (E.\,Routh, S.\,A.\,Chap\-lygin).}\label{pC}
In this case
\eq[2-14*]{
f_1=R, \qq f_2=R\gamma_3+a,
}
where~$R$ is the radius of ball, $a$ is the distance from the center of
mass to the geometrical center. The measure~$\rho $ is not constant any
more
\eq[eq2.14]{
\rho=\bigl( I_1I_3+I_1mR^2(1-\gamma_3^2)+I_3m(R\gamma_3+a)^2
\bigr)^{-\fracs12},
}
and the equations for~$K_1,\,K_2$ become trivial:~$\dot K_1=0$,
$\dot K_2=0$, i.\,e. the expressions
\eqc[eq2.15]{
K_1  = \om_3\rho^{-1} = \rho^{-1}\left(mR^2\gam_3({\bs M},\,{\bs\gam})+I_1M_3+
maR(({\bs M},\,{\bs\gam})+M_3\gam_3)+ma^2M_3\right)=\const,\\
K_2  =\frac{1}{R}({\bs M},\,{\bs r})=({\bs M},\,{\bs\gam})+
\frac{a}{R}M_3=\const.
}
are integrals of motion.

The integral~$K_2$ represents \emph{the Jellett integral}. This integral
is also present under the arbitrary law of friction at the point of
contact~\cite{markeev1}. The integral~$K_1$ was found by E.\,Routh in
1884~\cite{raus} and its form is a little bit mysterious. It was also
indicated by S.\,A.\,Chaplygin in the paper~\cite{chaplygin2}. Once again
we shall note that both integrals are linear with respect to the
velocities. They are the immediate generalizations of the cyclic integrals
corresponding to the precession~$\psi$ and to the proper
rotation~$\varphi$, but have no such natural dynamical
origin. The integral~$K_2$ sometimes is referred to as the
\emph{Chaplygin integral}.

\begin{rem*}
For axisymmetric bodies we can also indicate the other cases of existence
of simple quadratic integral of the form
$$
F=aK_1^2+bK_2^2,\quad a,b=\const.
$$
Obviously, we have to require in addition to condition~\eqref{eq2.1/2}
the following one
\begin{equation}
\label{rem-*1}
\frac{K_2^{-1}\dot K_1}{K_1^{-1}\dot K_2}=\frac{I_3\Bigl(1-\Bigl(\frac{f_2}{f_1}\Bigr)'\Bigr)}
{mf_1(f_1-f_2')}=\lambda=\const.
\end{equation}
The general solution of equations~\eqref{rem-*1} and~\eqref{eq2.1/2}
is expressed in hypergeometric functions. Among the axisymmetric figures
of the second order only the ball with displaced center satisfies these
equations. The closed bounded curves satisfying~\eqref{rem-*1}
and~\eqref{eq2.1/2} and different from ball are similar to the ovals.
Below we show that in this case the simple quadratic integral exists in
the totaly symmetric case.~$(I_1\ne I_2\ne I_3\ne
I_1)$.
\end{rem*}

\paragraph{Three-dimensional point maps in nonholonomic
mechanics.}\label{pD-new}
Before we consider the following cases of the body's motion, we shall
present some general construction that let us to establish relations
between equations~\eqref{eq1.1}, \eqref {eq1.2}, \eqref {eq1.3} to some point
one-to-one map in three-dimensional space. We present the computer analysis
of this map using the numerical integration of the indicated system at the
fixed value of energy. Using this method we can find out and give a visual
interpretation to various possibilities of existence of measure and
integrals in their various combinations.

To construct the three-dimensional map we use the Andoyer\f Deprit
variables $(L,G,H,l,g,h)$, which were regularly used in our
book~\cite{bormam-2} for computer (and analytical) research of Euler\f
Poisson, Kirchhoff and other Hamiltonian equations. As against to
nonholonomic situation, in the classical case these variables are
canonical, and by virtue of the fact that the area integral is always
present in the Euler\f Poisson type equations, we can limit ourselves in
the classical case to two-dimensional maps. The problems described above
require two additional integrals of motion; therefore, it is necessary to
use three-dimensional maps, and such maps are not necessarily possess an
invariant measure (as against to Hamiltonian mechanics). Using the known
formulas we make the transition from the variables $(\bM, \bs\gamma)$ to
the Andoyer\f Deprit variables\nopagebreak\vspace {-4mm}
\begin{align}
{}\span
M_1=\sqrt{G^2-L^2}\sin l,\quad M_2=\sqrt{G^2-L^2}\cos l ,\quad
M_3=L,\notag\\[-2pt]
\gamma_1&=\ts\left(\frac{H}{G}\sqrt{1{-}\left(\frac{L}{G}\right)^2}\!\!+
\frac{L}{G}\sqrt{1{-}\left(\frac{H}{G}\right)^2}\cos g\right)\sin l{+}
\sqrt{1{-}\left(\frac{H}{G}\right)^2}\sin g \cos l,\notag\\[-2pt]
\gamma_2&=\ts\left(\frac{H}{G}\sqrt{1{-}\left(\frac{L}{G}\right)^2}\!\!+
\frac{L}{G}\sqrt{1{-}\left(\frac{H}{G}\right)^2}\cos g\right)\cos l-
\sqrt{1{-}\left(\frac{H}{G}\right)^2}\sin g \sin l,\notag\\[-2pt]
\gamma_3&=\ts\left(\frac{H}{G}\right)\left(\frac{L}{G}\right){-}
\sqrt{1-\left(\frac{L}{G}\right)^2}
\sqrt{1-\left(\frac{H}{G}\right)^2}\cos g,\label{an-dp}
\end{align}
in which we can express the energy~$E=E(L,G,H,l,g)$.

In the Euler\f Poisson equations the value of~$H=(\bM, \bs\gamma)$ is
constant, but for equations~\eqref{eq1.1}, \eqref{eq1.2}, \eqref{eq1.3}
this is not the case any more. We fix the level of energy $E=E_0$, then
choose the intersecting plane, for example, as $g=g_0=\const$, and obtain
the three-dimensional map induced by sequential intersections of the phase
trajectory with the chosen intersecting plane. We present the map in the
variables $(L/G,H/G,l)$, because of its compactness by virtue of the fact
that $\Bigl|\frac LG\Bigr|\le 1$, $\Bigl|\frac HG\Bigr|\le 1$. Typical
examples of such three-dimensional maps are presented in
Figs.~\ref{ris-*1}, \ref{ris-*2}, \ref{ris-*3}, \ref {ris-*4}. It is
obvious that, because of the presence of one additional integral, the
trajectories are situated{\parfillskip=0pt}

\wfig<bb=0 0 56.9mm 55.3mm>{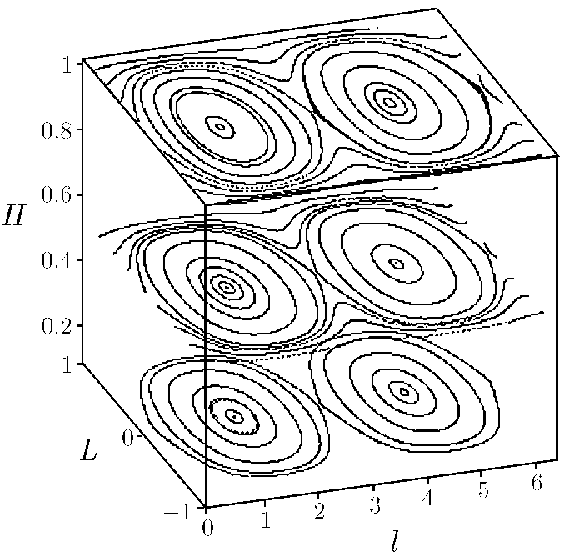}[ The three-dimensional map
described in subsection~\ref{pD-new} for the case of Chaplygin ball. The
figure shows very clearly that all trajectories are situated on joint
level surfaces of two integrals~$H=\const$ and $\bM^2=\const$ ($I_1=1$,
$I_2=2$, $I_3=3$) \label{ris-*1}]

\noindent on two-dimensional invariant manifolds of the point map, and the
presence of two additional integrals imply the stratification of the
three-dimensional space on invariant curves (Fig.~\ref{ris-*1}). In
general case when both the integrals and the measure are absent, the
complicated behavior of trajectories is possible. In this case random
motions alternate with the asymptotic attracting properties typical for
dissipative systems. Note also that as against to $(\bM, \bs\gamma)$ the
variables $L,G,H,l,g,h$ are more convenient for the analysis of
three-dimensional map, because in this variables the linear and angular
components are separated, and they have the obvious geometrical meaning
(see~\cite{bormam-2}).

One of examples of three-dimensional map is the well-known Smale\f
Williamson map. It does not preserve the measure, but is expressed by
analytical formulas. Other examples can be obtained by the study of
general (nonconservative) perturbations of two-degree Hamiltonian systems.

\paragraph{Rolling of balanced, dynamically nonsymmetric ball
(Chaplygin ball~\cite{chaplygin1}).}\label{pD} The equations of motion
of dynamically nonsymmetric ball with the center of mass coinciding with
the geometrical center can be written in the form\vspace{-3mm}
\eqc[eq2.16]{
 \dot{\bs M} = {\bs M}\times{\bs \omega}+
 \bs\gamma\times\pt{U}{\bs\gamma}, \qq
\dot{\bs\gamma}={\bs \gamma}\times{\bs\omega}, \\
 {\bs M}={\bf I}{\bs\omega}+D{\bs \gamma}\times({\bs\omega}\times{\bs\gamma}),
\qq D=ma^2,
}
where ${\bf I}=\diag(I_1,I_2,I_3)$ is the central tensor of inertia,
$U=U(\bs\gamma)$ is the potential energy. Equations~\eqref{eq2.16}
always have the measure with density $\rho$ and the first integrals of the
form
\eqc[eq2.17]{
 \rho=\frac{1}{\sqrt{1-D({\bs\gamma},\,{\bf A}{\bs \gamma})}}, \q
{\bf A}=({\bf I}+D{\bf E})^{-1}, \q {\bf E}=\|\dl_{ij}\|, \\
 H=\frac12({\bs M},\,{\bs \omega})+U(\bs\gamma),\q F_1={\bs \gamma}^2=1
 \q F_2=({\bs M},\,{\bs \gamma}).
}
At $U=0$ there exists the additional integral $F_3=\bM^2$ and the problem
become integrable (S.\,A.\,Chaplygin, 1903~\cite{chaplygin1});
the corresponding three-dimensional map is presented in Fig.~\ref{ris-*1}.

It was shown in the paper~\cite{kozlov1} that this problem is still
integrable with a Brun potential
\eq*{
U=\frac12k({\bf I}{\bs \gamma},\,{\bs \gamma}).
}
The integral $F_3$ in this case has the form
\eq*{
F_3=\bM^2- \frac{k}{\det{\bf A}}(\bs\gamma,{\bf A}\bs\gamma)
}

The authors indicated in~\cite{bm-new} that for any potential~$U$
the change of time $d\tau=\rho\,dt$ in equations~\eqref{eq2.16} makes them
Hamiltonian with a Poisson bracket, which is nonlinear with respect to the
phase variables~$(\bM, \bs\gamma)$ and has the form
\begin{equation}
\label{eq2b-*1}
\begin{gathered}
\{M_i,M_j\}=\eps_{ijk}\rho^{-1}(M_k-g\gamma_k),\quad \{M_i,\gamma_j\}=
\eps_{ijk}\rho^{-1}\gamma_k,\quad \{\gamma_i,\gamma_j\}=0,\\
g=D(\bs\omega,\bs\gamma)=\frac{D(\bs\gamma,{\bf A}\bM)}{1-D(\bs\gamma,
{\bf A}\bs\gamma)}.
\end{gathered}
\end{equation}
Bracket~\eqref{eq2b-*1} is degenerated; its Casimir functions are
integrals~$ F_1, F_2$~\eqref{eq2.17}. The Hamiltonian corresponding to
bracket~\eqref{eq2b-*1} is obtained from energy~\eqref {eq2.17} expressed
as a function of the moments by the formula
\begin{equation}
\label{eq2b-*2}
H=\frac12(\bM,{\bf A}\bM)+\frac12 g({\bf A}\bM,\bs\gamma)+U(\bs\gamma).
\end{equation}
After the change of variables $\bK=\rho\bM$ the Poisson bracket and the
Hamiltonian are represented in the form
\begin{equation}
\label{eq2b-*3}
\begin{gathered}
\{K_i,K_j\}=\eps_{ijk}(K_k-D\rho^2(\bK,\bs\gamma)a_k\gamma_k),\quad
\{K_i,\gamma_j\}=\eps_{ijk}\gamma_k,\quad \{\gamma_i,\gamma_j\}=0,\\
H=\frac12\rho^{-2}(\bK,{\bf A}\bK)+\frac12D({\bf A}\bK,\bs\gamma)^2+
U(\bs\gamma).
\end{gathered}
\end{equation}
Thus, at the zero level $(\bK, \bs\gamma)=0$ bracket~\eqref{eq2b-*3}
passes to the bracket described by algebra~$e(3)$. on which we can write
the Euler\f Poisson and Kirchhoff equations~\cite{bormam-2}.

Note that for the considered problem the density of measure $\rho$ is
the reducing multiplier (by Chaplygin~\cite{chaplygin3}). With its help the
nonholonomic equations are reduced to the Hamiltonian system. Chaplygin
himself used such reduction integrating the equations of motion of
nonsymmetric ball; as a preliminary he introduced a nonholonomic
analog of spherocon variables. It is possible to do these operations in
inverse order~\cite{chaplygin1}: one makes at first the change of
time $d\tau=\rho\,dt $ to receive a Hamiltonian system, and then
introduces the usual spherocon variables and use the Hamilton\f Jacobi
method.

As against to Poisson structure~\eqref{eq2-*3} related to the system
reduced on the field of a symmetry corresponding to proper rotation,
structure~\eqref{eq2a-*4} is related to complete system~\eqref{eq1.1},
\eqref{eq1.2}. Unfortunately, we were unable to generalize (to lift)
reduced structure~\eqref{eq2-*3} to such complete system. Possibly it is
either too difficult or some dynamic effects prevent such generalization.
Unfortunately, the dynamic effects preventing the reduction to Hamiltonian
form are very purely investigated~\cite{bormam-1}.

\paragraph{Rolling of unbalanced, dynamically nonsymmetric ball on plane.}
In this case equations \eqref{eq1.1}, \eqref{eq1.2} can be written in the
following convenient form
\begin{equation}
\label{eq-d*1}
\begin{gathered}
\left\{
\begin{aligned}
\dot {\bM}& = \bM\x \bs\omega+m\dot{\br}\x (\bs\omega\x \br),\\
\dot{\br}& = \br\x \bs\omega-\ba\x \bs\omega=(\br-\ba)\x\bs\omega
\end{aligned}
\right.\\
\bM={\bf I}\bs\omega+m\br\x(\bs\omega\x\br),
\end{gathered}
\end{equation}

\wfig<bb=0 0 38.7mm 30.0mm>{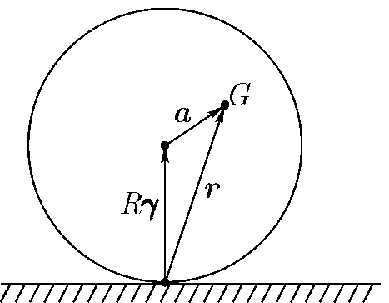}

\noindent where $\ba$ is the vector connecting the center of mass with the
geometrical center $\br=R\bs\gamma+\ba$ (see Fig.~\ref{ris6.eps}). It
turns out that in the case~$\ba\ne 0$ the integral $F_3=\bM^2$ of
system~\eqref{eq2.16} can be directly generalized and written in the form
\begin{equation}
\label{eq-d*2}
F=\bM^2-m\br^2(\bM,\bs\omega)=\bM^2-2m\br^2 H,
\end{equation}
where $H=\frac12(\bM,\bs\omega)$ is the energy integral. Though this
integral is simple enough, but probably it was not known earlier. Note
only that under the additional requirement of the dynamical symmetry the
Jellett integral and the Chaplygin integrals were found (see
subsection~\ref {pC}). Integral~\eqref{eq-d*2} can be considered as
their generalization for dynamically nonsymmetric situation.

\fig<bb=0 0 70.3mm 34.0mm>{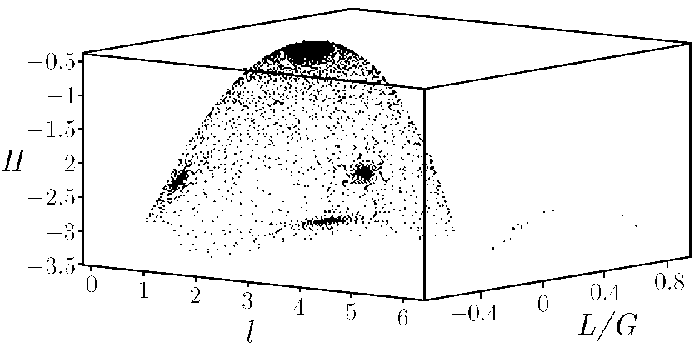}[One of trajectories in the problem
of rolling of unbalanced ball on plane. The figure shows clearly that all
points are situated on some surface; the condensations of points
correspond to asymptotic approximations of the trajectory to periodic
solutions. The trajectory goes out from the top and approaches to the three
points in lower part of surface\label{ris-*2}]

\fig<bb=0 0 118.4mm 50.0mm>{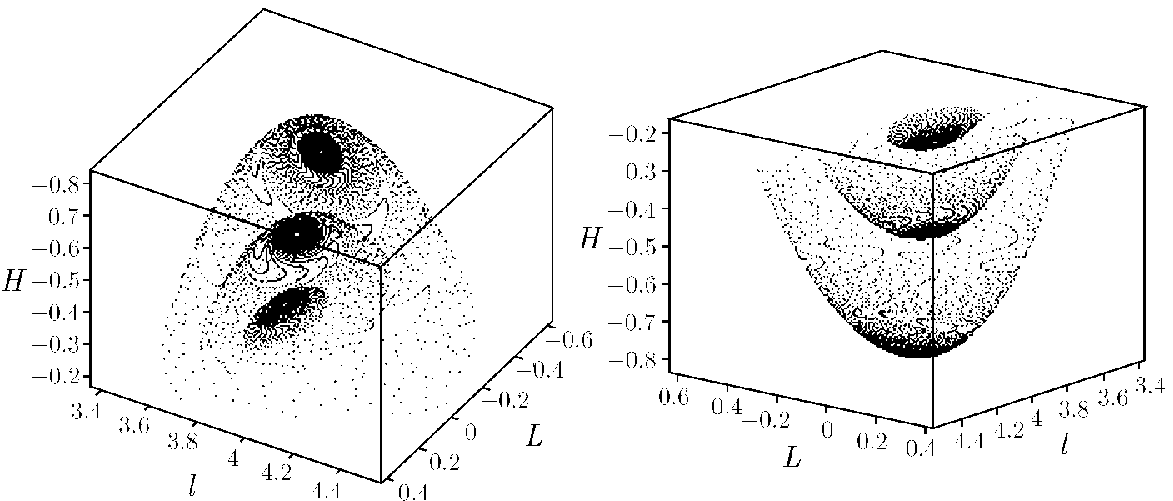}[Three trajectories in the
problem of rolling of unbalanced ball on plane. The figure shows very
clearly that points are situated on two-dimensional surfaces
(corresponding to the level of integral~\eqref{eq-d*2}). The condensation
of points corresponds to an asymptotic approximation to some periodic
solution.\label{ris-*3}]

We were unable to obtain generalizations of this integral for cases with
gyrostat and Brun field. Note also that for $\ba\ne 0$ there is probably
no measure. This is illustrated in Figs.~\ref{ris-*2}, \ref{ris-*3}.
These figures shows the asymptotic trajectories of the point
map situated on the two-dimensional surface of integral~\eqref{eq-d*2}.

\paragraph{An arbitrary body with a spherical central ellipsoid of
inertia.}\label{pE} $I_1=I_2=I_3=\mu $, $\mu=\const$.

This problem in general case requires both additional integrals,
nevertheless, there is always an invariant measure. For the first time
this fact was noted by V.\,A.\,Yaroschuk~\cite{yaroshuk}. In this case the
equations written in variables~${\bs \omega},{\bs \gamma} $ are more
convenient. They have form
\eqc[eq2.18]{
 (\mu+m{\bs r}^2)\dot{\bs \omega}  = m(\dot{\bs r}+
{\bs \omega}\times {\bs r})({\bs r},\,{\bs\omega}) -
m{\bs\omega}({\bs r},\,\dot{\bs r}) +
{\bs\gamma}\times\frac{\partial U}{\partial{\bs\gamma}}, \\
 \dot{\bs\gamma}  = {\bs\gamma}\times{\bs\omega}.
}

Equations~(\ref{eq2.18}) have an invariant
relation~$(\dot{\bs\omega},\,{\bs r})=0$, which is used for simplification
of some calculations.

The density of invariant measure for these equations indicated
in~\cite{yaroshuk} can be represented in the form
\eq[eq2.19]{
\rho=(\mu+{\bs r}^2)^{\fracs32}.
}
We were unable to reproduce the generalization of this measure for the case
of body's rolling on sphere indicated in the paper~\cite{yaroshuk}.
Probably, this generalization does not exist. Also it is not known,
what nontrivial integrable cases can be obtained with the help of
measure~(\ref{eq2.19}), and whether the system has any Hamiltonian origin,
possibly after the appropriate change of time.

\paragraph{Gyrostatic generalizations.}
Following mainly the paper by S.\,A.\,Chaplygin~\cite{chaplygin2}, we
present the generalizations of the indicated problems for the case with
additional uniformly rotating balanced rotor. The corresponding system can
be interpreted as a nonholonomic gyrostat. The gyrostatic effect can be also
obtained by an addition of multiply connected cavities completely filled
by the ideal incompressible liquid possessing nonzero circulation into the
body~\cite{bormam-2}. In the described case, equation for the
moment~\eqref{eq1.1} can be presented as
$$
\dot\bM=(\bM+\bS)\times \bs\omega+m\dot\br\times(\bs\omega\times\br)
+\bM_Q,
$$
where $S$ is the constant three-dimensional vector of gyrostatic moment.
It is easy to verify that the addition of rotor does not influence
the existence of invariant measure with the density depending on
the positional variables~$\bs\gamma$. %??????

{\bf a)} {\it Body of revolution.}
The equations of type~(\ref{eq2.5}) for the rotor with gyrostatic
moment~${\bs S}=(0,\,0,\,s)$ directed along the axis of revolution in
variables~\eqref{eq2.4} have the form
\eq[eq2.20]{
\frac{dK_1}{d\gamma_3} = -I_3\rho\Bigl(1-\Bigl(\frac{f_2}{f_1}\Bigr)'\Bigr)
K_2-s,\quad
\frac{dK_2}{d\gamma_3} = -m\rho f_1 ((f_1-f_2')K_1+f_2s).
}
Equations~(\ref{eq2.20}) were obtained in less convenient form by
S.\,A.\,Chaplygin~\cite{chaplygin2}. The density of invariant measure is
also defined by equation~\eqref{eq2.2}.

Let's consider sequentially the generalizations of the indicated earlier
cases of disk, ellipsoid, and ball with the displaced center.

{\bf b)} {\it Round disk.} Now equation~(\ref{eq2.9}) have to the following
form
\eqc[eq2.21]{
\frac{d^2\omega_3}{d\theta^2}-\ctg\theta\frac{d\omega_3}{d\theta} + mRI_3(R+a\ctg\theta)
\rho^2\omega_3=smR\rho^2(R+a\ctg\theta), \\
\rho=(I_1I_3+I_1mR^2+I_3ma^2)^{-\fracs12},
}
and at~$a=0$ in general case it is reduced to non-homogeneous
(for~$s\ne 0$) hypergeometric equation.

{\bf c)} {\it Ball with the displaced center of mass.}
Here system~(\ref{eq2.20}) has the form
\eq[eq2.24]{
\frac{dK_1}{d\gamma_3}  = -s,\quad
\frac{dK_2}{d\gamma_3} = -m\rho R(R\gam_3+a)s,
}
where $\rho$ is defined by relation \eqref{eq2.14}.
We can immediately show the first integral that generalizes the Jellett
integral
\eq[eq2.25]{
F=K_1+s\gamma_3=\const.
}
The second integral generalizing Routh (Chaplygin) integral has more
complicated nonalgebraic form
\begin{equation}
\label{eq2.26}
(I_1-I_3)\rho^{-1}\omega_3
- s\left\{\rho^{-1}-I_1\sqrt{\frac{ma^2}{I_1-I_3}}
\arctg\left( \sqrt{\frac{m}{I_1-I_3}}
\rho (R\gam_3(I_1-I_3)-aI_3)\right)\right\} = \const.
\end{equation}
In integral~(\ref{eq2.26}) we assume~$I_1>I_3$. For~$I_1<I_3$
the integral contain hyperbolic functions. Integral~(\ref{eq2.26})
was explicitly presented by A.\,S.\,Kuleshov~\cite{lit3}. It is
essentially simplified at~$I_1=I_3=\mu$ and has the form
\eq[eq2.27]{
\rho^{-1}\Bigl(3\mu\omega_3 - s
\frac{\mu+mR^2-2ma^2-maR\gamma_3}{ma^2}\Bigr)=\const.
}
The form of integral is even simpler for the case of balanced homogeneous
ball ($a=0$):
$$
\omega_3+\frac12\frac{mR^2\gamma_3^2}{\sqrt{\mu(\mu+mR^2)}}=\const.
$$

This simple integrable generalization was indicated by
D.\,K.\,Bobylev~\cite{bobylev} (some additional simplifications
in the case of explicit integration were also indicated by
N.\,E.\,Zhukovsky~\cite{Zhukovskij02}).

{\bf d)} {\it Dynamically nonsymmetric ball.}
The most general gyrostatic generalization for the case of Chaplygin ball was
suggested by A.\,P.\,Markeev~\cite{markeev}. The equations of motion and
the integrals are
\eqc[eq2-25]{
\dot{\bM}=(\bM+\bS)\times\bs\omega,\quad
\bs\gamma=\bs\gamma\times\bs\omega,\\
H=\frac12 (\bM,\bs\omega),\quad F_1=\bs\gamma^2=1,\quad
F_2=(\bM+\bS,\bM+\bS),\quad
F_3=(\bM+\bS,\bs\gamma),
}
where $\bS$ is the constant three-dimensional vector of gyrostatic moment.
Note that we were unable to generalize Poisson structure~\eqref{eq2b-*1}
to system~\eqref{eq2-25} for $\bS\ne 0$.

\paragraph{Rolling of ellipsoid on plane.}
It turns out that in the problem of rolling of balanced ellipsoid,
which axes are principal axes of inertia, there are cases of existence of
specific invariant measure, that are defined by restrictions on ratios of
moments of inertia and semiaxes of the ellipsoid of the surface. This
measure has found by V.\,A.\,Yaroschuk in~\cite{lit1}.

\fig<bb=0 0 111.8mm 57.4mm>{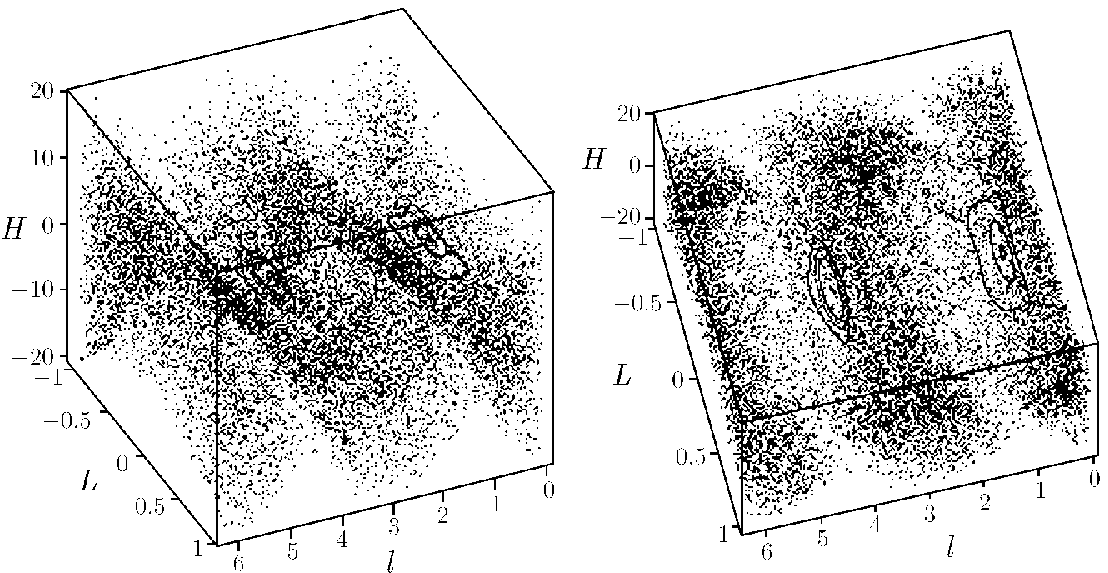}[Some trajectories in the
problem of rolling of ellipsoid with the spherical tensor of
inertia on plane. The figure shows clearly the regular trajectories filling
some curves that enclose stable periodic solutions (permanent rotations),
which in this case are degenerated as it was stated above.
A random layer (which is obtained from one trajectory) in this case is not
situated on any surface.\label{ris-*4}]

For the problem of ellipsoid's rolling, two cases of existence of the
invariant measure were already indicated in subsections~\ref{pD}
and~\ref{pE}. They are accordingly measures of balanced, dynamically
nonsymmetric ball \eqref{eq2.17} and of arbitrary body with the spherical
central ellipsoid of inertia~\eqref{eq2.19}. It is interesting that no
obstacles to existence of the analytical invariant measure of the general
rolling problem of balanced ellipsoid, which principal axes are principal
axes of inertia, are not found yet (as against to the case of Celtic
stone~\cite{kozlov1}). It is possible, that this measure exists (at least
in this situation there is no asymptotical behavior typical for the Celtic
stones), but is complicated and nonalgebraic.

For the surface of ellipsoid~$({\bs r},\,{\bf B}^{-1}{\bs r})=1$, where
${\bf B}=\diag(b_1,\,b_2,\,b_3)$, $b_i$ are squares of larger semiaxes,
we have the explicit expression
\eq[eq2.28]{
{\bs r}=\frac{{\bf B}{\bs\gamma}}{\sqrt{({\bf B}
{\bs\gamma},\,{\bs\gamma})}}.
}
If the cental tensor of inertia has the form
\eq[eq2.29]{
{\bf I}=\mu {\bf E}+\lambda m{\bf B},
}
then equations~(\ref{eq1.1}),\,(\ref{eq1.2}) have a measure only in the case
$\lambda=0$ and~$\lambda=1$ with the density (in variables~$\bM,\,\bs\gamma$)
\eq[eq2.30]{
\rho=(\mu+m{\bs r}^2)^{-\fracs12}=
\left((\mu+m{\bf B}){\bs r},\,{\bf B}^{-1}{\bs r}\right)^{-\fracs12}.
}

\begin{rem*}
For the case $\lambda=0$, measure~\eqref{eq2.30} was already
indicated in subsection~\ref{pE}. It is defined by formula~\eqref{eq2.19}
and is present for any surface of the body. The differences in powers
of expressions~\eqref{eq2.30} and~\eqref{eq2.19} are connected with the
various systems of variables~$(\bM, \bs\gamma)$ and $(\bs\omega,
\bs\gamma)$ and with the corresponding transformations of densities of
invariant measures.
\end{rem*}

Note that if equality~(\ref {eq2.29}) is fulfilled, then the motion
equations have a two-parameter set of vertical permanent rotations at
arbitrary real~$\lambda$ (in other cases this set is one-parameter).
A.\,V.\,Karapetyan in~\cite{lit2} showed that the conditions of existence
of such sets are even a little bit wider and have the form
\begin{equation}
\label{eq2-27}
\sum_{ijk}I_i(b_i-b_k)=0,
\end{equation}
where ${\bf I}=\diag(I_1,I_2,I_3)$. In addition to tensor~\eqref{eq2.29}
conditions~\eqref{eq2-27} are fulfilled in the case of nonholonomic
Chaplygin ball~($b_1=b_2=b_3=R^2 $) when there exist both the measure and
integral~\eqref {eq2.17}, and also the conditions hold in the axially
symmetric situation~$I_1=I_2$, $b_1=b_2$. Unexpectedly,
equality~\eqref{eq2.29} is also the necessary (but, generally speaking,
insufficient) condition of the integrability of equations of motion for the
case of ellipsoid on ideally smooth plane~\cite{burov}. Note the interesting
fact that for $\lambda\ne 0$ and~$\lambda\ne 1$, both the measure and
obstacles to its existence are not found. The three-dimensional section in
the case $\lambda=0$ is presented on Fig.~\ref {ris-*4}.

All the results described above that are connected to the rolling of body
on plane are presented in table~1.

\begin{table}[!htp]
\unitlength=1mm
\begin{picture}(0,0)
\put(49.8,132.5){\cfig{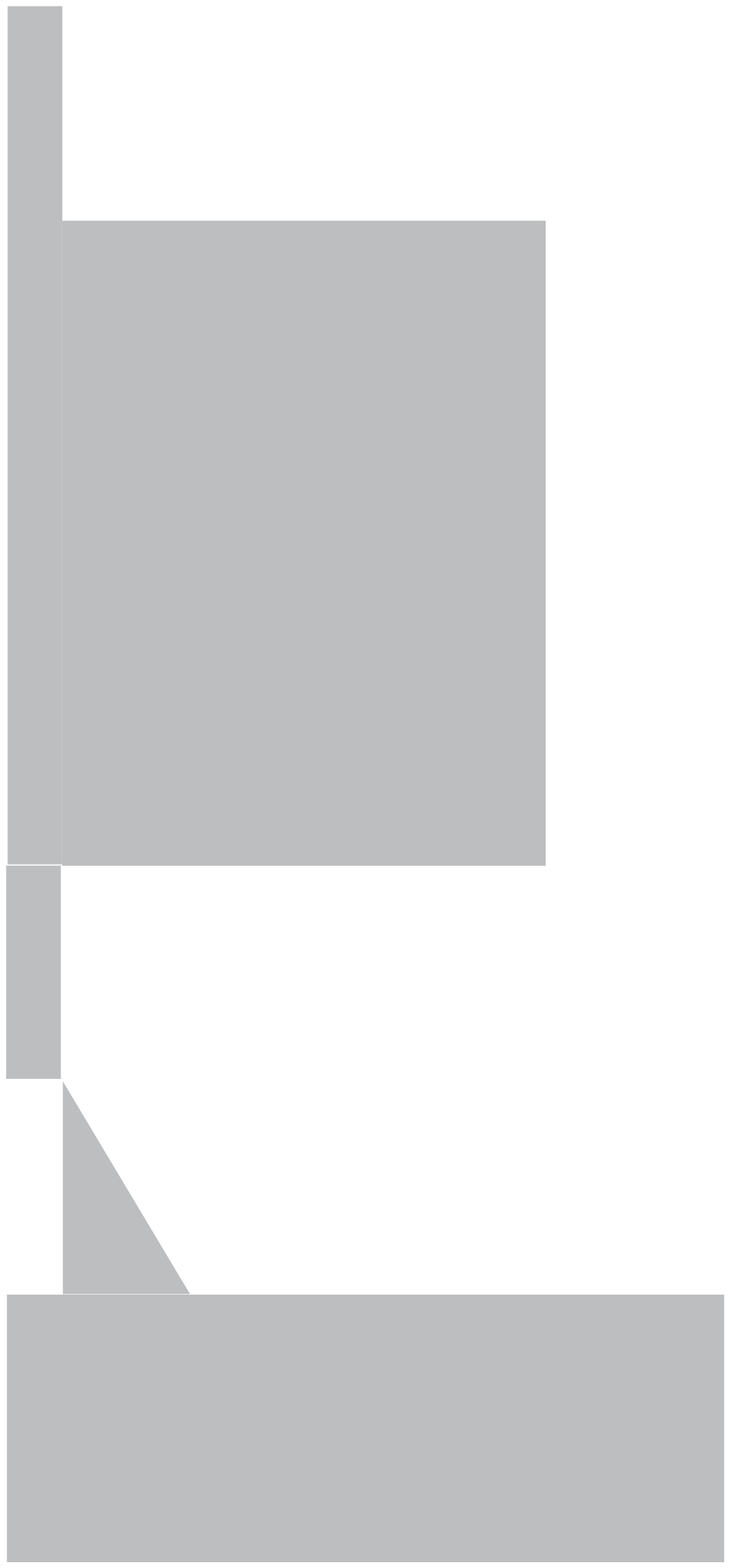}}
\end{picture}
\resizebox{!}{\textheight}{\rotatebox{90}{\parbox{1.05\textheight}{\footnotesize\vspace{5mm}
\begin{center}
\mbox{}\hfill Table 1. Rolling of body on plane\\[2mm]
\tabcolsep=2pt
\begin{tabular}{|l||c|c|c|c|c|c|c|}
\hline
\parbox{2,3cm}{\raggedright tensor of inertia}&\multicolumn{3}{|c|}
{\parbox{6cm}{\centering dynamically nonsymmetric case\\
                        $I_1\neq I_2\neq I_3\neq I_1$}} &
\multicolumn{3}{|c|}{\parbox{6cm}{\centering axial dynamical symmetry\\
                        $I_1=I_2,\,U=U(\gamma_3)$}} &
\parbox{3cm}{\centering total dynamical symmetry\\ $I_1=I_2=I_3=\mu$} \\
\hhline{|-||-|-|-|-|-|-|-|}
\parbox{2.4cm}{surface\\ of body}  &
\multicolumn{2}{|c|}{ball} &
\parbox{30mm}{\centering ellipsoid} &
\parbox{30mm}{\centering an arbitrary body of revolution} &
\parbox{30mm}{\centering round disk with sharp edge} &
\parbox{30mm}{\centering unbalanced ball} &
\parbox{30mm}{\centering arbitrary} \\
\hhline{|-||-|-|-|-|-|-|-|}
\parbox{2.4cm}{geometrical and dynamical restrictions} &
\parbox{33mm}{\centering the center of mass coincides with the geometrical center} &
\parbox{24mm}{\centering the center of mass does not coincide with the geometrical center} &
\parbox{30mm}{\centering the axes of dynamical and geometrical ellipsoid
coincide ${\bf I}=\mu {\bf E}+m{\bf B}$}&
\multicolumn{3}{|c|}{\parbox{6,5cm}{\centering
the geometrical and dynamical axes coincide and contain the center of
mass}} &
\parbox{30mm}{\centering --- }\\
\hhline{|=::=:=:=:=:=:=:=|}
\parbox{2.4cm}{measure}  &
\parbox{25mm}{\centering $(1{-}D(\bs\gam,{\bf A}\bs\gam))^{-\fracs12}$} &
unknown &
\parbox{30mm}{\centering $(\mu+m\br^2)^{-\fracs12}$}  &
\parbox{30mm}{\centering $(I_1I_3+m(\br,{\bf I}\br))^{-\fracs12}$} &
\parbox{30mm}{\centering $\const$} &
\parbox{30mm}{\centering $(I_1I_3+m(\br,{\bf I}\br))^{-\fracs12}$}  &
\parbox{30mm}{\centering $(\mu +m\br^2)^{\fracs32}$ (for variables
$\bs\omega$ and $\bs\gam$)}\\
\hhline{|-||-|-|-|-|-|-|-|}
\parbox{2.4cm}{additional integrals}  &
\parbox{27mm}{\centering $\arr{\bM^2=\const\\ (\bM,\bs\gam)=\const}$\\
(two integrals)} &
\parbox{30mm}{\centering $\arr{\bM^2-m\br^2(\bM,\bs\omega)=\\
=\const}$\\ (one integral)} &
 \parbox{30mm}{\centering none of integrals are found} &
 \parbox{30mm}{\centering two integrals are obtained from the solution
 of system of two linear equations~\eqref{eq2.5}} &
 \parbox{30mm}{\centering two integrals are obtained from the solution of
 hypergeometric equation~\eqref{eq2.9}}&
 \parbox{30mm}{\centering $\arr{\omega_3/\rho=\const\\(\bM,\,\br)=\const}$}
 & \parbox{30mm}{\centering none of integrals are found} \\
\hhline{|-||-|-|-|-|-|-|-|}
\parbox{2.4cm}{integrable\\ addition\\ of gyrostat}   &
\parbox{30mm}{\centering possible\\ (A.\,P.\,Markeev\\ 1986)} &
\parbox{20mm}{\centering it seems to be \\ impossible} &
\parbox{30mm}{\centering the measure is preserved} &
\parbox{30mm}{\centering S.\,A.\,Chaplygin (1897)} &
\parbox{30mm}{\centering S.\,A.\,Chaplygin (1897)} &
\parbox{30mm}{\centering at $I_1=I_2=I_3$ the gyrostat was added by
  D.\,K.\,Bobylev,\\ at $I_1=I_2\ne I_3$ by A.\,S.\,Kuleshov (2000)} &
  \parbox{30mm}{\centering the measure is preserved} \\
\hhline{|-||-|-|-|-|-|-|-|}
\parbox{2.4cm}{Hamiltonian form} &
\parbox{34mm}{\centering the system is Hamiltonian after the change of time (A.\,V.\,Borisov,
I.\,S.\,Mamaev, 2001)}&
\parbox{25mm}{it seems that the system is not Hamiltonian}&
\parbox{25mm}{\centering unknown} &
  \multicolumn{3}{|c|}{\parbox{6,5cm}{\centering the reduced system
  is Hamiltonian after the change of time, defined by the reducing multiplier
  \\ (A.\,V.\,Borisov, I.\,S.\,Mamaev, 2001)}}&
\parbox{30mm}{\centering unknown}  \\
\hhline{|-||-|-|-|-|-|-|-|}
\parbox{2.4cm}{authors} &
\parbox{20mm}{\centering S.\,A.\,Chaplygin (1903)} &
\parbox{20mm}{\centering A.\,V.\,Borisov,\\
 I.\,S.\,Mamaev (2001)} &
\parbox{30mm}{\centering V.\,A.\,Yaroschuk (1995)} &
\parbox{30mm}{\centering S.\,A.\,Chaplygin (1897)}&
\parbox{30mm}{\centering S.\,A.\,Chaplygin (1897), P.\,Appell,
  D.\,Korteweg (1898)} &
\parbox{30mm}{\centering E. J. Routh (1884),
  S.\,A.\,Chaplygin (1897)} &
  \parbox{30mm}{\centering V.\,A.\,Yaroschuk (1992.)} \\
\hhline{|-||-|-|-|-|-|-|-|}
\parbox{2.4cm}{generalizations\\ and remarks} &
\parbox{38mm}{\centering the integrable addition of Brun field is possible\\ (V.\,V.\,Kozlov, 1985)\\
the Hamiltonian form is preserved for arbitrary fields with the loss of one integral}&
\parbox{25mm}{the Brun field can not be added (preserving the integral)}&
--- & --- & ---& --- & ---\\
\hline
\end{tabular}\\
\end{center}
{\small Remark. The cases of existence of the corresponding (tensor) invariants are
indicated by gray color in the table. The partial filling corresponds to
the uncomplete set of integrals.}}}}
\end{table}

\section{Body on a Sphere}

Now we consider systematically the situations analogous to one in the case
of plane that originate in the problem of rolling of body on a sphere.
First of all we shall note that kinematic equation~(\ref{eq1.3})
can be written as
\eq[eq3.1]{
\dot{\bs\gam}=\bs\gam\x\bs\om\mp k\dot {\bs r}, \qq k=1/R_0,
}
where the "minus"\ sign means the rolling on interior of surface of sphere
(Fig.~\ref{ris2.eps}\,a), and the~"plus"\ sign means the rolling on
exterior of surface of sphere (Fig.~\ref{ris2.eps}\,b).

\fig<bb=0 0 86.5mm 60.5mm>{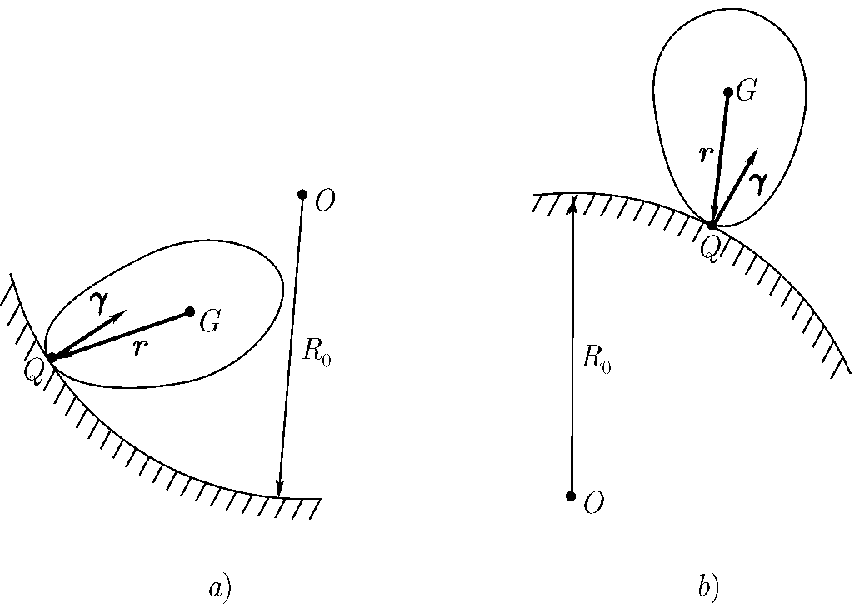}

\paragraph{Rolling of body of revolution.}
The body of revolution is defined analogously by formulas~(\ref{eq2.1}),
and we can obtain the explicit form of the density of invariant measure
\eqc[eq3.2]{
\rho(\gam_3) =\rho_0(1-kf_1)^3 (1-k f_2'),\\
\rho_0 = \bigl(I_1I_3+m(\bs r,{\bf I}\br)\bigr)^{-1/2}=
\bigl(I_1I_3+mI_1f_1^2(1-\gamma_3^2)+mI_3f_2^2\gamma_3^2\bigr)^{-1/2}
}
and establish the presence of the symmetry field~$\bv$ with
operator~\eqref{eq2.3}.\goodbreak

Reduced system (\ref{eq2.5}) in the variables of type~\eqref{eq2.4}
\eq[eq3.2-1]{
K_1=(\bM,\br)(1-kf_1)f_1^{-1},\quad  K_2=\omega_3\rho_0^{-1},\quad
}
has the form
\eqc[eq3.3]{
\frac{1}{\rho_0}\frac{dK_1}{d\gamma_3}  = -\biggl[I_3\biggl(1-\Bigl(\frac{f_2}{f_1}\Bigr)'
\biggr)+kf_1(I_1-I_3)\bigl(1-kf_2'\bigr) \biggr] K_2\,,\\
\frac{1-kf_1}{\rho_0}\frac{dK_2}{d\gamma_3}  =
- mf_1\bigl(f_1-f_2'-kf_1f_2'\bigr)K_1+
mk\rho_0I_1f_1^2 \left(\gam_3f_2'+
\sqrt{1-\gam_3^2} \Bigl(f_1\sqrt{1-\gam_3^2}\,\Bigr)' \right)K_2\,.
}
Equations~\eqref{eq3.3} in somewhat different variables connected with
the semifixed axes were obtained by P.\,V.\,Woronetz~\cite{voronec},
who generalized the Chaplygin arguments to the case of sphere.

Integrating the equations of system~(\ref{eq3.3}) we determine the
dependence on time of the nutation angle~$\ta=\arccos(\gam_3)$ using the
quadrature of the energy integral
$$
\begin{gathered}
H=\frac12\frac{I_1+m\br^2}{1-\gamma_3^2}\dot\gamma_3^2+\\
+\frac{1}{2I_1(1-\gamma_3^2)}\Bigl(\frac{K_1^2}{(1-kf_1)^2}-
\frac{I_3}{mf_1^2}K_2^2+\frac{mf_2^2}{I_1}\Bigl(
\frac{K_1}{1-kf_1}-\frac{K_2}{m\rho f_1f_2}\Bigr)^2\Bigr)+U(\gamma_3).
\end{gathered}
$$

For the problem of rolling of the body of revolution on sphere we can
describe the reduced system in variables~\eqref{eq2.4} and the
corresponding Poisson structure similarly to the case of plane. We do
not present the corresponding calculations because of their bulkiness.

Similarly to the problem of rolling on a plane we shall consider
some special cases.

\paragraph{The problem of rolling of round disk} with the center of
masses displaced along the axis of symmetry in general case is not reduced
to the hypergeometric equation any more, but the system nevertheless
becomes simpler; thus, the measure is not constant any more. Same as
above, $f_1, f_2 $ are defined by relations (\ref{eq2.71}), and the
density of invariant measure has the following form
\eq*{
\rho=\rho_0\left(1-\frac{kR}{\sqrt{1-\gam_3^2}}\right)^3,\q
\rho_0=\Bigl(I_1 I_3 + m\bigl(I_1R^2+I_3a^2\bigr)\Bigr)^{-1/2}=\const,
}
where $R$ is the disk radius and $a$ is the displacement of the center of
mass.

The second-order equation has the form
\eq[eq3.4]{
\frac{d^2\om_3}{d\ta^2}+\ctg\ta
\left(1+\frac{kR}{\sin\ta-kR}\right)\om_3
-\frac{\rho_0^2mRI_3}{1-\frac{kR}{\sin\ta}}
\left(R+a\ctg\ta+\frac{kR^2(I_1-I_3)}{I_3\sin\ta}\right)\om_3=0,
}
where $\gam_3=\cos\ta$.\goodbreak

This equation in the particular case~$I_3=2I_1$, $a=0$ (the
homogeneous balanced disk) was obtained by P.\,Woronetz~\cite{voronetz3},
and at~$a=0$ it was investigated in~\cite{kholm0} by methods of the
qualitative analysis. In particular, the stability of stationary motions
and the probability of falling of disk on sphere was investigated
and this probability is found to be equal to zero. In the
paper~\cite{kholm} it was also shown that, as against to the nonholonomic
problem, the the Hamilton system describing the motion of disk on a sphere
with absolute (ideal) sliding that seems to be simpler is not integrable
any more, and its behavior has random properties. The number of degrees of
freedom for this system is increased in comparison with the case of plane,
where the indicated system is integrable because of the preservation
of impulse's horizontal component, and this is the reason for such result.

\paragraph{Ball with displaced center.}
Here functions $f_1,f_2$ are also defined by relations (\ref{2-14*}), and
the expression for measure~$\rho$ is the same as in the case of plane:
\eq*{
\rho=\Bigl(I_1I_3+I_1mR^2(1-\gam_3^2)
+I_3m(R\gam_3+a)^2\Bigr)^{-1/2}.
}
In variables~\eqref{eq3.2-1} equations~(\ref{eq3.3}) have the form
\eq[eq3.7]{
\frac{1}{\rho}K_1'=-kR(I_1-I_3)(1-kR)K_2,\q
\frac{1}{\rho}K_2'=\frac{kmR^3}{1-kR}K_1.
}
With the help of these equations we obtain two linear with respect to
$K_1,K_2$ nonalgebraic integrals of the form
\eq[eq3.8]{
F_2=\Bigl(\sqrt{m(I_3-I_1)}\,K_2+\frac{mR}{1-kR}\,K_1\Bigr)
e^{\lambda\tau},\quad
F_3=\Bigl(\sqrt{m(I_3-I_1)}\,K_2-\frac{mR}{1-kR}\,K_1\Bigr)
e^{-\lambda\tau},
}
where $\lambda^2=mk^2R^4(I_3-I_1)$, $\tau=\int\rho_0(\gamma_3)\,d\gamma_3$,
and the additional quadratic algebraic integral (dependent on $F_2$,$F_3$)
\eq[eq3-8]{
F=F_2F_3=\frac{mR^2}{(1-kR)^2}K_1^2+(I_1-I_3)K_2^2.
}
The integrals $F_2$, $F_3$ are new and generalize Routh and Jellett
integrals~\eqref{eq2.15}. Integral~\eqref{eq3-8} was found by
A.\,S.\,Kuleshov~\cite{lit3}. In the case of spherical tensor of
inertia~$I_3=I_1$ we have
\eqc[eq3.9]{
K_1=(\bM,\,\br)=M_1\gam_1+M_2\gam_2+ M_3\Bigl(\gam_3+\frac{a}{R}\Bigr) =
\const,\\
(1-kR)K_2-kmR^3K_1\int\rho(\gamma_3)\,d\gamma_3=\const,
}
i.\,e. $K_1$ coincide with the classical Jellett integral.

\wfig{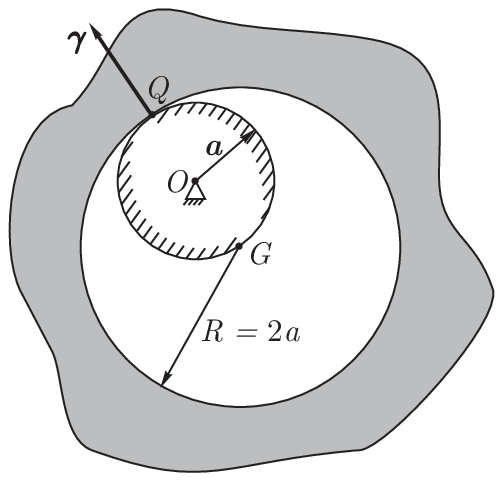}

\paragraph{Balanced, dynamically nonsymmetric ball on sphere.}
We shortly describe an integrable case connected with the problem of
rolling of balanced, dynamically nonsymmetric ball on sphere analogous to
the motion of Chaplygin ball on plane (see \eqref{eq2.16},
\eqref{eq2.17}). This system is defined by the equations
\eqc[eq3.12]{
\dot {\bM}  = \bM\times\bs\om,\quad \dot{\bs\gam}  = k\bs\gam\times\bs\om,\\
\bM=({\bf I}+D)\bs\om-D\bs\gam(\bs\om,\bs\gam),
}
where~$k=\frac{R}{R-a}$, $a$ is the radius of ball, $m$ is its mass,
$D=ma^2$, $R$ is the radius of the fixed sphere (see Fig.~\ref{ris3.eps}).
(In the case of plane $R\to \infty$ and~$k=1$.)
Equations~\eqref{eq3.12} always have the integrals
\eq*{
H=\frac12(\bM,\bs\om),\q F_1=\bs\gam^2=1, \q F_2=\bM^2.
}

\noindent
and the invariant measure with density~$\rho$~\eqref{eq2.17}. We need one
more integral for the complete integrability of system (such as the area
integral at $k=1 $). It exists only under the additional condition~$a=2R$
found by A.\,V.\,Borisov in~\cite{BF}, which corresponds to the rolling of
balanced, dynamically nonsymmetric ball on the interior of the fixed
sphere (see Fig.~\ref{ris3.eps}). This integral have the form
\eq[eq3.13]{
F=(\bM,\ol{\bf A}\bs\gam), \qq \ol{\bf A}={\bf E}-2(\tr({\bf I}
+D))^{-1}({\bf I}+D)
}
and can be generalized to case~\eqref{eq3.12} with the additional
Brun potential~$U=\frac12({\bf I}\bs\gam,\bs\gam)$~\cite{BF}. Using the
transformations~$\wt{\bM}=\ol{\bf A}\bM$, $\wt{\bs\gamma}=\bs\gamma$ under
condition~$a=2R$ we transform equations~\eqref{eq3.12} into the equations
describing the motion of Chaplygin ball on the horizontal plane.
For arbitrary parameter~$k$ and potential~$U$ the Poisson structure of
equations~\eqref{eq3.12} similar to one indicated at~$k=1$ in
subsection~2\,\ref {pD} (formula~\eqref{eq2b-*1}) is still unknown.

\paragraph{Unbalanced, dynamically nonsymmetric ball on sphere.}\label{pD*}
In this case there exists one (and only one) quadratic integral
generalizing the corresponding result on plane. The equations are
\begin{equation}
\label{eq-d*-1}
\left\{
\begin{aligned}
\dot{\bM}& = \bM\x\bs\omega+m\dot{\br}\x(\bs\omega\x\br),\\
\dot\br& = \frac{1}{1-kR}(\br-\ba)\x \bs\omega,
\end{aligned}
\right.
\end{equation}

\wfig<bb=0 0 47.1mm 62.1mm>{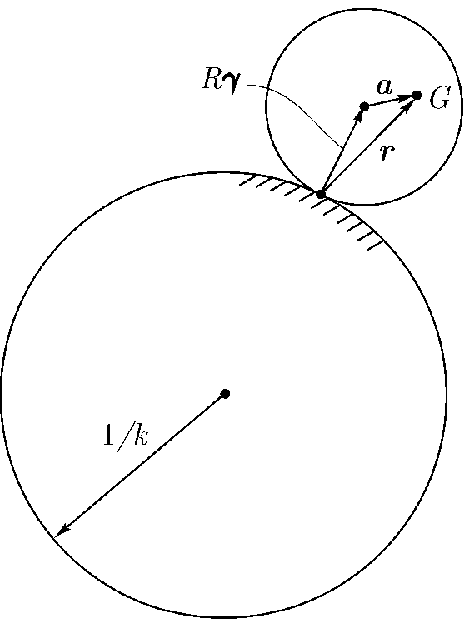}

\noindent
where $\br=R\bs\gamma+\ba$, $\bM={\bf I}\bs\omega+m\br\x(\bs\omega\x\br)$.
It has exactly the same form
\begin{equation}
\label{eq-d*-2}
F=\bM^2-m\br^2(\bM,\bs\omega).
\end{equation}
Note also that for~$a\ne 0$ it seems that there is no invariant
measure, and also generalizations for the cases with additional
gyrostat or with the field of Brun problem are unknown. In the case of
rotational symmetry $I_1=I_2$, integral~\eqref{eq-d*-2} was indicated by
A.\,S.\,Kuleshov and the acquaintance with this result has induced authors
to the analysis of dynamically nonsymmetric situation.

\paragraph{Gyrostatic generalizations.}
Let's briefly discuss the integrable gyrostatic generalizations. So
to preserve the integrability for the body of revolution's case we should
direct the balanced rotor with the moment~$\bS$ along the axis of dynamical
symmetry. In variables~\eqref{eq3.2-1} the analog of system~(\ref {eq2.20})
has the following form
\eqc[eq3.10]{
\frac{1}{\rho_0}K_1'=-\left[I_3\left(1-\left(\frac{f_2}{f_1}\right)'\right)
+kf_1(I_1-I_3)(1-kf_2') \right] K_2 + \\
+s\rho_0^{-1}(1-kf_2')(1-kf_1),\\
\frac{1-kf_1}{\rho_0}K_2'=km\rho_0I_1f_1^2
\left[\gam_3f_2'+\sqrt{1-\gam_3^2}\Bigl(f_1\sqrt{1-\gam_3^2}\Bigr)'
\right]K_2-\\
-mf_1\bigl(f_1-f_2'-kf_1f_2'\bigr)K_1- smf_1f_2(1-kf_2')(1-kf_1),
}
where $\rho_0$ is defined by relation \eqref{eq3.2}; thus, the linear
system is non-homogeneous.

In the case of round disk, the integrals and equations~(\ref{eq3.10})
can not be essentially simplified. In the case of {\it ball with the
displaced center},~$f_1,f_2$ are defined by equations~\eqref{eq2.14}, and
equations~\eqref{eq3.10} are
\eqc[eq3.11]{
\frac{1}{\rho_0}K_1' = -kR(I_1-I_3)(1-kR)K_2-\frac{1}{\rho_0}R(1-kR)^2s,\\
\frac{1}{\rho_0}K_2' = \frac{mR^3k}{1-kR}K_1 - mR(1-kR)(R\gam_3+a)s.
}
The first integrals in this case can not be written with the help of
elementary functions.

In the case of total dynamical symmetry~$I_1=I_3$ with the help of
equations~\eqref{eq3.11} we obtain the explicit integral of Jellett type
$$
F_2=K_1+(1-kR)^2\Bigl(\gamma_3+\frac{a}{R}\Bigr)=
\bigl(\bM R^{-1}+\bS(1-kR),\br\bigr)(1-kR)=\const.
$$
Expressing from this integral $K_1$ and substituting the expression in the
first equation of system~\eqref{eq3.11}, we obtain the explicit quadrature
for $K_1(\gamma_3)$. The gyrostatic generalizations of
integral~\eqref{eq3.13} for the case of nonsymmetric ball are unknown.

\wfig{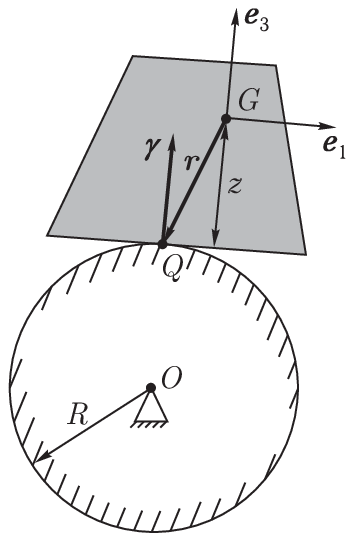}

\paragraph{Rolling of body with partially flat surface on sphere.}
Let's consider one more problem connected with the rolling of body on a
sphere, which has no analog in the case of plane. We consider the rolling
of the body with the flat foundation on the exterior surface of sphere
(Fig.~\ref{ris4.eps}). This problem for the first time was considered by
P.\,Woronetz, who indicated the integrability of the problem in the case
of rotational symmetry. Let's  write the equations of motion in the frame
of references connected with the body in the case when the field of force
is absent:
\eq[eq3.14]{
\dot {\bM}  = \bM\x\bs\om+m\dot{\br}\x(\bs\om\x{\br}), \q
\bM  = {\bf I}\bs\om+m\bs\gam\x(\bs\om\x\bs\gam),
}
where~$\bf I$ is the central tensor of inertia, $m$ is the mass of body.

For the flat part of surface we have~$\br=(r_1,\,r_2,\,z)$,
$z=\const$, $\bs\gam=(0,\,0,\,1)$,
where~$r_1,\,r_2$ are the projections of position of the center of mass on
the flat foundation (see Fig.~\ref{ris4.eps}), and using the fact
that~$\dot{\bs\gam}=0$ we obtain for them using equation~\eqref{eq1.3}
\eq[eq3.15]{
\dot r_1=k^{-1}\om_2, \qq \dot r_2=-k^{-1}\om_1, \qq k=R^{-1}.
}
We can specify two cases, when equations~(\ref{eq3.14}),\,(\ref{eq3.15})
have invariant measure.

\mbox{}{\bf a)} $z=0$~--- {\it the center of mass is situated on the
contact plane}, and tensor of inertia~${\bf I}=\diag(I_1,\,I_2,\,I_3)$
is arbitrary. For the variables~$\bM,\br$ the density of invariant
measure is
\eq[eq3.16]{
\rho(\bM,r_1,r_2)=\Bigl(I_3+m\br^2\Bigr)^{-1/2}.
}
It can be also written for the equations in variables~$\bs\om,\,r_1,\,r_2$:
\eq[eq3.17]{
\rho(\bs\omega,r_1,r_2)=(I_1I_2+m(\br,{\bf I}\br))(I_3+m\br^2)^{\fracs12}.
}

\begin{rem*}
In this form under additional and not essential restriction~$I_3=I_1+I_2$
(i.\,e. in the case of flat plate) the invariant measure was indicated by
V.\,A.\,Yaroschuk~\cite{yaroshuk}.
\end{rem*}

In the considered case in addition to the measure, equations~\eqref{eq3.14},
\eqref{eq3.15} have one (and only one for~$I_1\ne I_2$) first integral
\begin{equation}
\label{eq*}
F=\bM^2-2m\br^2H,\quad H=\frac{1}{2}(\bM,\bs\omega),
\end{equation}
independent of energy integral. It was not known earlier.

Besides, it turns out that integral~\eqref{eq*} \emph{is transferred
without modifications to the case~$z\ne 0$, for which the measure is not
known} (and probably does not exist). Note that we already see the similar
situation in the problem of rolling of {\em unbalanced}, dynamically
nonsymmetric  ball (see subsection~\ref{pD*}). The lack of measure
for~$z\ne 0$ obviously prevent the existence of Hamiltonian form.
For~$z\ne 0$ though the measure exists, the system is probably also not
Hamiltonian even after the appropriate change of time.

{\bf b)} $I_1=I_2$, $z\ne 0$, {\it i.\,e. the center of mass is situated
on the axis of dynamical symmetry}. In variables~$(\bM,\br)$
the density of invariant measure is
\eq[eq3.18]{
\rho  = (I_1I_3+m(\br,{\bf I}\br))^{-\fracs12}.
}
This system is already integrable. Indeed, in variables
$$
\begin{gathered}
K_1=(\bM,\br)=M_1r_1+M_2r_2+M_3r_3,\\
K_2=\frac{\omega_3}{\rho}=\frac{mz(M_1r_1+M_2r_2)+(I_1+mz^2)M_3}
{\sqrt{I_1I_3+m(\br,{\bf I}\br)}}
\end{gathered}
$$
the equations of motion of system have linear form
\eq[eq3.19]{
\frac{dK_1}{du}=\frac{I_1-I_3}{2R\sqrt{I_3(I_1+mz^2)+I_1mu}}K_2,\quad
\frac{dK_2}{du}=\frac{m}{2R\sqrt{I_3(I_1+mz^2)+I_1mu}}K_1.
}
P.\,Woronetz in the paper~\cite{voronetz1} noted that their explicit
solution can be obtained in elementary functions. This solution
defines two linear with respect to~$\bM $ additional first integrals.

There are two different methods of solving of~\eqref{eq3.19}.

\begin{itemize}\itemsep=-1pt
\item[1)] {\boldmath $I_1<I_3$.}
\begin{equation}
\label{eq-**}
\begin{gathered}
K_1=\sqrt{I_1-I_3}\bigl(-c_1\cos\vfi(u)+c_2\sin\vfi(u)\bigr),\q
K_2=\sqrt{m}\bigl(c_1\sin\vfi(u)+c_2\cos\vfi(u)\bigr),\\
\vfi^2=\frac{(I_1-I_3)(I_1I_3+I_3mz^2+I_1mu)}{I_1^2mR^2},\quad
c_1,c_2=\const.
\end{gathered}
\end{equation}
\item[2)] {\boldmath $I_1>I_3$,}
In this case we should use hyperbolical instead of trigonometrical
functions
\end{itemize}

In this case there is also simple, but dependent quadratic integral
\begin{equation}
\label{eq-***}
F=mK_1^2+(I_1-I_3)K_2^2,
\end{equation}
that evidently is a particular case of~\eqref{eq*}.
The energy integral can be presented in the form
\begin{equation}
\label{eq-***3}
\begin{gathered}
H=\frac12(\bM,\bs\omega)+U(\br)=\frac18\frac{I_1+mz^2+mu}{R^2u}\dot u^2+\\
+\frac{1}{2I_1^2u}\Bigl(
(I_1+mz^2)K_1^2+(I_1u+I_3z^2)K_2^2-\frac{z}{I_1^2u}
\sqrt{I_1I_3+I_3mz^2+I_1mu}K_1K_2\Bigr)+U(u),
\end{gathered}
\end{equation}
and using this form we obtain after the integration of
system~\eqref{eq3.19} the explicit quadrature for~$\dot u$.

In the axisymmetric case we can add rotor with the moment~$\bS=(0,0,s)$
along the axis of dynamical symmetry; measure~\eqref{eq3.18} is
preserved, and equations~(\ref{eq3.19}) are
\eq[eq3.21]{
2R\frac{dK_1}{du}=\frac{I_1-I_3}
{\sqrt{I_3(I_1+mz^2)+I_1mu}}K_2-s,\quad
2R\frac{dK_2}{du}=-\frac{m(K_1+zs)}{\sqrt{I_3(I_1+mz^2)+I_1mu}}.
}
The general solution of system~\eqref{eq3.21} is presented as a
superposition of the solution of homogeneous equation at~$s=0$
\eqref{eq-**} and the particular solution of non-homogeneous
equation~\eqref{eq3.21} of the following form
$$
K_1^{(part)}=\frac{sz(I_3+I_1(1+R/z))}{I_1-I_3},\quad
K_2^{(part)}=\frac{s\sqrt{I_3(I_1+mz^2)+I_1mu}}{I_1-I_3}.
$$
Thus quadratic integral~\eqref{eq-***} in this case is presented as
$$
F=m(K_1-K_1^{(part)})^2+(I_1-I_3)(K_2-K_2^{(part)})^2.
$$

\begin{rem*}
We have noted in the paper two nontrivial cases of non-existence of
one quadratic integral. The similar integral is present in the case of
homogeneous ball's motion on the triaxial ellipsoid. The origin of such
integrals is connected with the presence of similar integrals in the
axisymmetric situation, when the system is completely integrable, and there
are two linear integrals. However, in the general case the dependence on
the positional variables in these integrals is complicated and is
expressed in special functions. In three cases that we have found, the
linear integrals are expressed in elementary functions and generate the
quadratic integral, which is represented by rational function.
This integral can be also generalized to the nonsymmetric situation.
\end{rem*}

\section{Conclusions}
In this paper we collect all the known at present cases of existence of
invariant measure, integrals, Poisson structure for the equations of
nonholonomic rolling of rigid body on plane and sphere. In all conceivable
assortment of situations we have not found out any case when {\em there are
two integrals, but no measure}. Possibly it is connected with the
specificity of the equations of nonholonomic mechanics.

Depending on the presence of this or that set of invariants there are
qualitative distinctions in the behavior of system. The system can exhibit
both typically Hamiltonian properties and the properties of conservative
systems realized on the example of Celtic stones' problem. For the analysis
of the indicated situations we use the method of {\em three-dimensional
Poincar\'{e} maps} and exactly this method at first let us to find out
integrals~\eqref{eq-d*2}, \eqref{eq-d*-2}, \eqref{eq*} numerically,
and then to obtain their explicit form. Undoubtedly, the research of
three-dimensional point maps in the cases of presence and lack of the
measure both from analytical and from the computational point of view
allows to find out many remarkable effects in nonholonomic systems.

From the point of view of this approach the problem of global evolution
of the Celtic stone~\cite{lit2,markeev2} is the most interesting. In this
problem we have to study the invariant and asymptotic manifolds on the
level surface of three-dimensional maps (not preserving an area) realizing
under various restrictions on the parameters of system.

Authors thank A.\,V.\,Karapetyan for useful discussions and A.\,A.\,Kilin
for the help with realization of numerical experiments.

\begin{table}[!p]
\begin{center}
\unitlength=1mm
\begin{picture}(0,0)
\put(48.5,132.4){\cfig{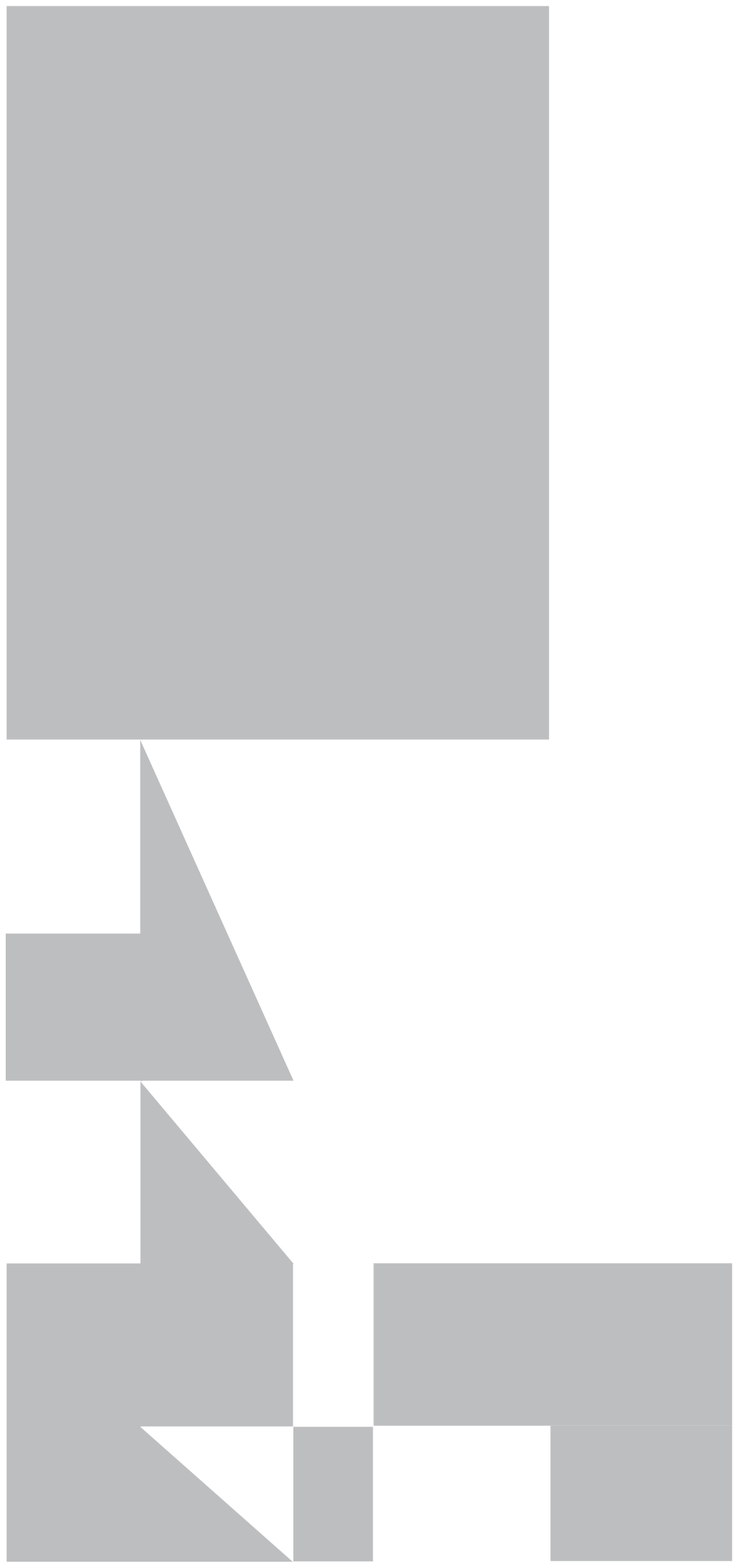}}
\end{picture}
\resizebox{!}{\textheight}{\rotatebox{90}{\parbox{1.7\textwidth}{\footnotesize
\tabcolsep=2pt
\def\arraystretch{2}
\mbox{}\hfill {\small Table 2. Rolling of the body on sphere}\\[3mm]
\medskip
\begin{tabular}{|l||c|c|c|c|c|c|c|c|c|}
\hline
\parbox{2,2cm}{\raggedright tensor of inertia}
& \multicolumn{5}{|c|}{dynamically nonsymmetric case $I_1\ne I_2\ne I_3\ne I_1$} &
\multicolumn{4}{|c|}{dynamical symmetry $I_1=I_2$, $U=U(\gamma_3)$} \\
\hhline{|-||--|-|-|-|-|-|-|-|}
\parbox{24mm}{surface\\ of the body} &
\multicolumn{3}{|c|}{ball} &
\multicolumn{2}{|c|}{body with partially flat surface} &
\parbox{30mm}{\centering an arbitrary body of revolution} &
\parbox{28mm}{\centering round disk} &
\parbox{30mm}{\centering unbalanced ball} &
\parbox{30mm}{\centering body with partially flat surface} \\
\hhline{|-||-|-|-|-|-|-|-|-|-|}
\parbox{24mm}{geometrical and dynamical restrictions} &
\multicolumn{2}{|c|}{\parbox{50mm}{\centering the center of mass
    coincides with the geometrical center}} &
\parbox{20mm}{\centering the center of mass does not
    coincide with the geometrical center} &
\parbox{20mm}{\centering  the center of mass is situated on the contact plane $(z=0)$} &
\parbox{15mm}{\centering  the center of mass is not situated on the contact plane} &
\multicolumn{4}{|c|}{\parbox{100mm}{\centering geometrical and dynamical
axes coincide and contain the center of mass}} \\
\hhline{|~||-|-|~|~|~~~~~|}
 & & $a=2R$ & & & & \multicolumn{4}{|c|}{}\\
\hhline{|=::=:=:=:=:=:=:=:=:=|}
measure
& \multicolumn{2}{|c|}{\parbox{50mm}{\centering $(1-D(\bs\gam,\,{\bf
A}\bs\gam))^{-1/2}$\\(V.\,A.\,Yaroschuk, 1992)}}
& unknown
& \parbox{24mm}{\centering $(I_3+m\br^2)^{-1/2}$\\ (V.\,A.\,Yaroschuk, 1992,\\
A.\,V.\,Borisov, I.\,S.\,Mamaev, 2001)}
& unknown
& \multicolumn{3}{|c|}{\parbox{60mm}{\centering $\frac{(1-kf_1)^3(1-k(\gamma_3f_2)')}
{\sqrt{I_1I_3+m(\br,{\bf I}\br)}}$\\ (A.\,V.\,Borisov, I.\,S.\,Mamaev, 2001)}} &
$(I_1I_3+m(\br,{\bf I}\br))^{-1/2}$\\
\hhline{|-||-|-|-|--|-|-|-|-|}
integrals &
\parbox{20mm}{\centering $\bM^2=\const$\\one integral}
& \parbox{24mm}{\centering $\bM^2=\const$,\\ $(\bM,\,\ol{\bf A}\bs\gam)=\const$\\ (two integrals)\\
(A.\,V.\,Borisov, 1994)}
& \parbox{30mm}{\centering $\bM^2-m\br^2(\bM,\bs\omega)=\const$\\ one integral\\
(A.\,V.\,Borisov, I.\,S.\,Mamaev, 2001)}
& \multicolumn{2}{|c|}{\parbox{50mm}{\centering $\bM^2-m\br^2(\bM,\bs\omega)=\const$\\
(A.\,V.\,Borisov, I.\,S.\,Mamaev, 2001)}}
& \parbox{30mm}{\centering
two integrals are obtained from the solution of system of two linear
equations~\eqref{eq3.3}\\ (P.\,V.\,Woronetz, 1909)} &
\parbox{30mm}{\centering
two integrals are obtained from the solution of second order
equation~\eqref{eq3.4}\\ (P.\,V.\,Woronetz, 1909)}
& \parbox{30mm}{\centering two integrals are expressed in elementary
functions~\eqref{eq3.8}\\ (A.\,S.\,Kuleshov, 2000,
A.\,V.\,Borisov, I.\,S.\,Mamaev, 2001)}
& \parbox{30mm}{\centering two integrals are expressed in elementary
functions~\eqref{eq-**}\\ (P.\,V.\,Woronetz, 1911)}\\
\hhline{|-||-|-|-|-|-|-|-|-|-|}
\parbox{24mm}{integrable\\ addition\\ of gyrostat} &
\parbox{22mm}{\centering possible without loss of integral and measure} &
\parbox{12mm}{\centering not found} &
\parbox{12mm}{\centering not found} &
\parbox{18mm}{\centering not found} &
\parbox{15mm}{\centering not found}
& \multicolumn{4}{|c|}{\parbox{115mm}{\centering
possible along the axis of dynamical symmetry}}\\
\hhline{|-||-|-|-|-|-|-|-|-|-|}
\parbox{25mm}{Hamiltonian form}
& \parbox{22mm}{\centering probably, the system is not Hamiltonian}
& \parbox{20mm}{\centering the system is Hamiltonian after the change of time
(A.\,V.\,Borisov, I.\,S.\,Mamaev, 2000)}
& \parbox{22mm}{\centering probably, the system is not Hamiltonian}
&  \multicolumn{2}{|c|}{probably, the system is not Hamiltonian} &
\multicolumn{4}{|c|}{\parbox{120mm}{\centering the reduced system is Hamiltonian
after the change of time defined by the reducing multiplier
(A.\,V.\,Borisov, I.\,S.\,Mamaev, 2001)}}\\
\hhline{|-||-|-|-|-|-|-|-|-|-|}
\parbox{24mm}{generalizations\\ and remarks}
& \parbox{22mm}{\centering the addition of Brun field is possible
preserving one integral and measure}&
\parbox{22mm}{\centering the integrable addition of Brun field is possible
(A.\,V.\,Borisov, Yu.\,N.\,Fedorov (1994))}
& \parbox{24mm}{\centering the Brun field could not be added
(preserving the integral)} &
\parbox{24mm}{\centering V.\,A.\,Yaroschuk found the measure under
additional nonessential restriction $I_3=I_1+I_2$} &
--- & --- & --- &
\parbox{24mm}{\centering A.\,S.\,Kuleshov found only one quadratic
integral dependent on two present linear integrals}&
\parbox{26mm}{\centering P.\,V.\,Woronetz found the solution in
quadratures under additional nonessential restriction
$I_1=I_2=\frac{1}{2}I_3$, $z=0$}\\
\hline
\end{tabular}\\
\bigskip
{\small Remark. The cases of existence of the corresponding (tensor) invariants are
indicated by gray color in the table. The partial filling corresponds to
the uncomplete set of integrals.}
}}}
\end{center}
\end{table}

\end{document}